\documentstyle[11pt,epsf,fbsty]{article}
\epsfverbosetrue
 
\font\tenrm=cmr10

 1
 1
 1

\def\npb#1#2#3{    {\it Nucl. Phys. }{\bf B\,#1} (19#2) #3}
\def\plb#1#2#3{    {\it Phys. Lett. }{\bf B\,#1} (19#2) #3}
\def\prd#1#2#3{    {\it Phys. Rev. }{\bf D\,#1} (19#2) #3}
\def\prep#1#2#3{   {\it Phys. Rep. }{\bf #1} (19#2) #3}
\def\prl#1#2#3{    {\it Phys. Rev. Lett. }{\bf #1} (19#2) #3}

\def\zpc#1#2#3{    {\it Zeit. f\"ur Physik }{\bf C\,#1} (19#2) #3}

\def\hph#1#2#3{    {hep--ph/{#1}{#2}{#3}}}

\def\bbbone{{\mathchoice {\rm 1\mskip-4mu l} {\rm 1\mskip-4mu l}
{\rm 1\mskip-4.5mu l} {\rm 1\mskip-5mu l}}}

\def\ltap{\ \raisebox{-.4ex}{\rlap{$\sim$}} \raisebox{.4ex}{$<$}\ }
\def\gtap{\ \raisebox{-.4ex}{\rlap{$\sim$}} \raisebox{.4ex}{$>$}\ }
\def\gsim{{~\raise.15em\hbox{$>$}\kern-.85em
          \lower.35em\hbox{$\sim$}~}}
\def\lsim{{~\raise.15em\hbox{$<$}\kern-.85em
          \lower.35em\hbox{$\sim$}~}}
\def\nhalf{\mbox{$\frac{1}{2}$}}

\def\beq{\begin{equation}}
\def\eeq{\end{equation}}
\def\bea{\begin{eqnarray}}
\def\eea{\end{eqnarray}}

\def\spd#1{{\partial_{#1}}}
\newcommand{\nn}{\nonumber}
\newcommand{\wti}{\widetilde}
\newcommand{\ti}{\tilde}

\def\tanb{\mbox{$\tan \! \beta\,$}}

\def\ggg{{(g^2 \!+\!g_Y^2)}}
\def\epsp{{\epsilon_+}}
\def\epsm{{\epsilon_-}}
\def\duno{\partial_{\epsilon_d}}
\def\ddue{\partial_{\epsilon_u}}
\def\dtre{\partial_{\epsilon_3}}
\def\diii{\partial_{\epsilon_i}}

\newcommand{\bsg}{\mbox{$b \rightarrow s \gamma\,$}}
\newcommand{\br}{\mbox{$Br( b \rightarrow s \gamma)\,$}}

\def\al{\alpha}
\def\wal{\mbox{$\wti \alpha$}}

\newcommand{\vev}{\rm{VEV}\,}
\newcommand{\Lm}{\Lambda}
\newcommand{\Gm}{\Gamma}
\begin{document}
\begin{titlepage}
\noindent
\phantom{a}     \hfill         WIS-96/50/Dec.-PH    \\[1.01ex]
\phantom{a}     \hfill         December 1996        \\[7ex]

\begin{center}

{\bf On the  Minimal Messenger Model}         \\[11ex]
{\bf    Francesca M.\ Borzumati     }                     \\[1ex]
{\it Department of Particle Physics, Weizmann Institute, Rehovot, 
               Israel                                    } \\[20ex]
\end{center}
{\begin{center} ABSTRACT \end{center}}
\vspace*{1mm}

\parbox{14.4cm}
{
\noindent
We study the Minimal Messenger Model, a minimal version of Gauge 
Mediated Supersymmetry Breaking models. Boundary conditions equal to 
zero for trilinear and bilinear soft parameters at the messenger scale 
make this model free from the supersymmetric CP problem and extremely 
predictive. These boundary conditions and the vicinity of the messenger 
scale to the electroweak one, requires a careful implementation of the 
mechanism of radiative breaking of $SU(2)\times U(1)$. We assess the 
importance of considering the complete one--loop effective potential 
and of including a set of logarithmic two--loop corrections to the $B$ 
parameter for the correct determination of the electroweak minimum. We 
analyze the resulting low--energy spectrum and give predictions of 
interest for future experimental searches. 
}
\vfill
\end{titlepage}

\thispagestyle{empty}
\phantom{aa}
\newpage

\setcounter{page}{1}

%
\section{Introduction}

Models with low--energy breaking of supersymmetry, communicated by 
gauge interactions to the observable sector, have recently drawn 
considerable attention as interesting alternatives to models in which 
this communication is mediated by gravity~\cite{DNNS}. The minimal 
realization of the latter at the electroweak scale is the well known 
Minimal Supersymmetric Standard Model (MSSM).

The former, the so--called gauge mediated supersymmetry breaking 
(GMSB) models have, indeed, several attractive features. Most 
important of these is the fact 
that, as gauge interactions are flavor blind, squark and slepton 
masses are universal. Given the low value of the supersymmetry 
breaking scale, this universality is hardly broken by the evolution of 
mass parameters to the electroweak scale through Renormalization
Group Equations (RGE). (Universality means, in this context, that 
scalar masses are only functions of gauge quantum numbers, and that 
$A$--terms are small or proportional to fermion yukawa couplings.) 
Moreover, they can be more predictive than the MSSM, having a smaller
number of free parameters, and may, at least in a minimal 
variant~\cite{BABU}, provide a solution to the supersymmetric 
CP problem~\cite{DNS}.

In this minimal version, dubbed the Minimal Messenger Model (MMM), 
trilinear and bilinear soft parameters vanish at the messenger scale
$X$. Moreover, if $X$ is of ${\cal O}(\Lm)$, with $\Lm$ the ratio of 
the messenger F-term over $X$, after the radiative breaking of the 
electroweak gauge symmetry is implemented, this model turns out to be 
practically a one--parameter model, $\Lm$. 

Although the MMM was already considered in~\cite{BABU,DIM,BAGG}, the 
mechanism of breaking of the electroweak gauge group was not always 
correctly implemented. Furthermore, not all experimental constraints 
on the model were always included. The aim of this paper is to present
a comprehensive analysis addressing these issues. After a definition 
of the model in Sect.~2, we will impose the breaking of 
$SU(2)\times U(1)$ through minimization of the RGE improved 
tree--level Higgs potential (Sect.~3.1). In Sect.~3.2, we will 
demonstrate the importance of considering the full one--loop 
corrected effective 
potential for the determination of the electroweak minimum. We will 
also argue on the need to include a set of additional two--loop
corrections to the parameter $B$ of same size than those induced by 
the one--loop effective potential. Finally, we will discuss the 
viability of the model for different values of $\Lm$. In the last 
section, Sect.~4, we will verify which regions of $\Lm$ survive the 
imposing of experimental bounds coming from direct searches of 
supersymmetric particles and the indirect constraint due to the 
measurement of $b \to s \gamma$ and list the main predictions of this 
model.

\section{The model}

The messenger sector of the MMM consists of only one pair of chiral 
superfields, $\Phi$, $\bar \Phi$, which transforms as a vectorlike
representation of the electroweak gauge group. In particular, having in 
mind an embedding of this model in an underlying Grand Unified Theory 
(GUT), $\Phi$, $\bar \Phi$ are chosen to be in a $5$, $\bar 5$ 
representation of $SU(5) \supset SU(3)_C\times SU(2)_L \times U(1)_Y$. 
They couple at the tree--level to a singlet $S$ 
($W = \lambda S \Phi \bar \Phi$). For simplicity, we neglect the 
difference in the evolution of $\lambda$ for the components of the 
two 5--plets with different $SU(3)_C$ and $SU(2)_L$ quantum numbers. 

The scalar component of $S$ acquires a vacuum expectation value (\vev)
giving therefore a 
supersymmetric mass $\lambda <\!\!S\!\!>$ to fermion and scalar 
components of both messengers. The auxiliary component of $S$ 
acquires a \vev as well, and gives rise to a supersymmetry--violating 
mass term, $\lambda F_S$, which mixes the scalar components of $\Phi$ 
and $\bar \Phi$. The information of 
supersymmetry breaking is then transmitted to the visible sector
via gauge interactions. A soft 
mass is generated for the i--th gaugino ($i = 1,2,3$) at the 
one--loop level, with fermion and scalar components of $\Phi$ and 
$\bar \Phi$ exchanged as virtual particles. This is:
\beq 
 M_i\,(\lambda <\!\!S\!\!>)  \ = \  
 \frac{1}{4 \pi} \, \alpha_i (\lambda <\!\!S\!\!>) \, 
                    \frac{F_S}{<\!\!S\!\!>} \, g(x)\,; 
 \qquad x \equiv \frac{F_S}{\lambda <\!\!S\!\!>^2} \,,
\label{bcgaug0}
\eeq
where a GUT normalization for the $U(1)_Y$ coupling was chosen:
$\alpha_1 = (5/3) \alpha_Y = (5/3) \alpha / \cos^2\theta_W$. If we 
indicate with 
$\Lm$ the ratio $F_S/<\!\!S\!\!>$ and with $X$ the supersymmetric 
mass $\lambda <\!\!S\!\!>$, we can rewrite (\ref{bcgaug0}) in a 
more compact form as: 
\beq 
 M_i(X)  =  \wal_i(X) \,\Lm \, g(x)\,; \qquad \quad 
            \wal_i(X) \equiv \frac{\alpha_i(X)}{(4 \pi)}\,.
\label{bcgaug}
\eeq
A tilde on gauge and yukawa couplings indicates hereafter a 
division over $4 \pi$. Natural values for $x $ are of ${\cal O}(1)$ 
(i.e. $(\lambda <\!\!S\!\!>)^2 \sim \lambda F_S$), but at $x=1$ 
massless scalar messengers appear in the theory. We choose in the
following $x=1/2$ and we shall comment on modifications obtained 
in the limit $x \to 1$. In the range $0-1$ the function $g(x)$, 
given in~\cite{SMARTIN}, is monotonically increasing and has 
values $g(0) = 1$, $g(1)= 1.386$. 

Communication of supersymmetry breaking is passed to the scalars of
the observable sector at the two--loop level with scalar and fermion 
components of the messenger fields, gauge bosons and gauginos
exchanged as virtual particles. The masses obtained are: 
\beq
 {m_i^2(X)}  =    2 \Lm^2   \left\{ 
     C_3 \,\wal_3^2(X) + C_2 \,\wal_2^2(X) 
           + \frac{3}{5} Y^2 \,\wal_1^2(X)
                            \right\}\,f(x)    \,,
\label{bcscal}       
\eeq
where $C_3 = 4/3, 0$ for triplets and singlets of $SU(3)_C$, 
$C_2 =3/4,0$ for doublets and singlets of $SU(2)_L$; $Y =Q-T_3$ is the 
hypercharge and $i$ runs over all scalars present in the 
theory. The function $f(x)$, derived in~\cite{DGP} 
and~\cite{SMARTIN}, is almost always a flat 
function equal to 1, except for a sharp drop at $x=1$, where it has 
the value 0.7.  

Finally, trilinear and bilinear couplings in the soft scalar 
potential~\footnote{This relation, the explicit form of the 
superpotential 
$ W  =     h^{ij}_e  H_d L_i E^c_j 
       +   h^{ij}_d  H_d Q_i D^c_j 
       -   h^{ij}_u  H_u Q_i U^c_j 
       - \mu^\Lm H_d H_u  
$,
and the multiplication rule $ H_d Q \equiv \epsilon_{ij} {H_d}_i Q_j$; 
$ \epsilon_{12} = -\epsilon_{21} = 1$,
define the sign--conventions used in this analysis} 
\beq
 V_{soft} \ni
 - (A_u h_u)^{ij} H_u {\wti Q}_i {\wti U}^c_j  + 
   (A_d h_d)^{ij} H_d {\wti Q}_i {\wti D}^c_j  + 
   (A_e h_e)^{ij} H_d {\wti L}_i {\wti E}^c_j  
 - B \mu \, H_d H_u 
\label{vsoft}
\eeq
(where ${\wti Q}$, ${\wti U}^c$, ${\wti D}^c$ etc., as well as 
 $H_d$ and $H_u$ indicate here the 
scalar components of the corresponding superfields) 
vanish at the messenger scale: 
\beq
 A_u^{ij}(X) = A_d^{ij}(X) = A_e^{ij}(X) = 0\,; \quad \quad \quad 
      B(X)   = 0\,. 
\label{bcaabb}
\eeq
The last relation in (\ref{bcaabb}) is the identifying property of 
this model. (For a discussion on how such a boundary condition can 
be theoretically implemented, see~\cite{DNS}.) Phenomenologically, 
it renders the model very predictive; technically, it makes the 
search of the correct electroweak minimum rather complicated.

Low--scale ($M_Z$) inputs of our analysis are: $ \alpha_3 =0.120$, 
$\alpha_2= 0.0335$, $\alpha_1=0.0168 $, corresponding to 
$\alpha^{-1} = 127.9 $ and $ \sin^2\theta_W = 0.2316$, and loosely 
compatible with a gauge couplings unification~\cite{NIR}. As running 
fermion masses we use $m_t(M_Z) = 171\,$GeV, $m_b(M_Z)=3.0\,$GeV and 
$m_\tau(M_Z)= 1.75\,$GeV.

\section{Radiative Breaking of $SU(2)_L\times U(1)_Y$}
\subsection{RGE improved tree--level Higgs potential}

We evolve all the initial parameters (\ref{bcgaug}),
(\ref{bcscal}), and (\ref{bcaabb}), which depend only on $\Lm$, 
from $X$ ($= 2 \Lm$) to a decoupling scale $Q_0$.
Henceforth, we shall refer to $X$ as to the
``high--scale'', as compared to the ``low--scale'' $\sim M_Z$. 
The evolution is performed using the one--loop MSSM RGE. 
For reference, we report in Appendix~A 
these equations and the two--loop equation for $B$, 
in the approximation of Kobayashi--Maskawa matrix $K \sim \bbbone$. 
Those actually used for this analysis, with all intergenerational 
mixing terms needed for the calculation of $b \to s \gamma$, can be 
found in~\cite{BBMR}. 

Given the much more modest evolution of the masses of weakly 
interacting 
sparticles, and the fact that they are much lighter than the strongly 
interacting ones, we take $Q_0^2$ to be the geometrical mean of the 
high--scale values of $m^2_{\wti Q}$ and $m^2_{\wti U}$. These depend 
on the decoupling scale itself (see (\ref{bcscal})) since the 
high--scale gauge couplings are to be obtained from our inputs at 
$M_Z$ through a running of the Standard Model (SM) RGE from $M_Z$ to 
$Q_0$ and the MSSM RGE from $Q_0$ to $X$. A simple iteration allows 
to find this scale rather quickly. For 
$\Lm \sim 100\,$TeV, $Q_0$ is typically $\sim 1\,$TeV. All squark 
masses cluster around this value. Corrections for the inadequacy of 
this scale for the weak mass parameters involved in the breaking 
mechanism will be provided by the inclusion of the one--loop 
corrections to the scalar potential, leaving therefore an overall 
scale ambiguity of the next order. Unless a different value is 
explicitly mentioned, the choice of $Q_0$ specified above is that
made throughout this paper. It will appear obvious later on why this
is indeed a good choice. 

For the high--scale yukawa couplings required as inputs of the MSSM 
RGE, we need the value of $\tanb$. This parameter, together with 
$\mu$, is obtained by imposing that the electroweak minimum is a 
minimum of the neutral Higgs potential, and, indeed, the deepest one.
We postpone the discussion of the latter point to a later moment.
As for the former, we require that the first derivatives with respect 
to the neutral higgses $H_u^0$, $H_d^0$ of the RGE improved 
tree--level potential 
\beq
 V_0\,(Q_0)  = 
    \mu^2_{H_d} \left\vert H^0_d\right\vert^2 
  + \mu^2_{H_u} \left\vert H^0_u\right\vert^2
  - (B \mu)     \left(H^0_d H^0_u+ h.c.\right) + \frac{\ggg}{8}
     \left(\vert H^0_d\vert^u-\vert H^0_u\vert^2\right)^2
\label{higgspot}
\eeq
( $ \mu^2_{H_i} \equiv m^2_{H_i} + \mu^2$ ($i=u,d$)) are equal to 
zero when it is $\langle H_u^0 \rangle = v_u$ and 
$\langle H_d^0 \rangle = v_d$: 
\bea
 \left( 
 m^2_{H_d} + \mu^2 + \nhalf\,\ggg\,(v_d^2-v_u^2)
 \right) 2 v_d  - (B\mu)\, 2 v_u                       & = & 0
\label{derivtreeone}                                     \\
 \left( 
 m^2_{H_u} + \mu^2 - \nhalf\,\ggg\,(v_d^2-v_u^2)
 \right) 2 v_u  - (B\mu)\, 2 v_d                       & = &  0\,,   
\label{derivtreetwo}
\eea
and that the obtained solution is a minimum. This minimization 
condition has to be imposed at the low--energy scale $M_Z$. Gauge 
couplings are indeed evolved down to $M_Z$, whereas, as said before, 
the evolution of mass parameters is stopped at $Q_0$: hence, the 
$Q_0$--dependence of the potential $V_0$.    
The equations (\ref{derivtreetwo}) are more often cast in the form:
\beq
 \mu^2   =  
 \frac{{m_{H_d}^2 -m_{H_u}^2 \tan^2\beta}}{{\tan^2\beta-1}}
            - \frac{1}{2} M_Z^2\,;   
\quad \quad \quad \quad 
  \sin 2\beta  =  
 \frac{2 B\mu}{{m^2_{H_u} +m^2_{H_d} +2 \mu^2}} \,.
\label{mincond}
\eeq
Self--consistent solutions of this problem are obtained through 
a numerical iteration. 

\begin{figure}[t]
\begin{center} 
\epsfxsize=16.0 cm
\leavevmode
\epsfbox[72 495 556 715]{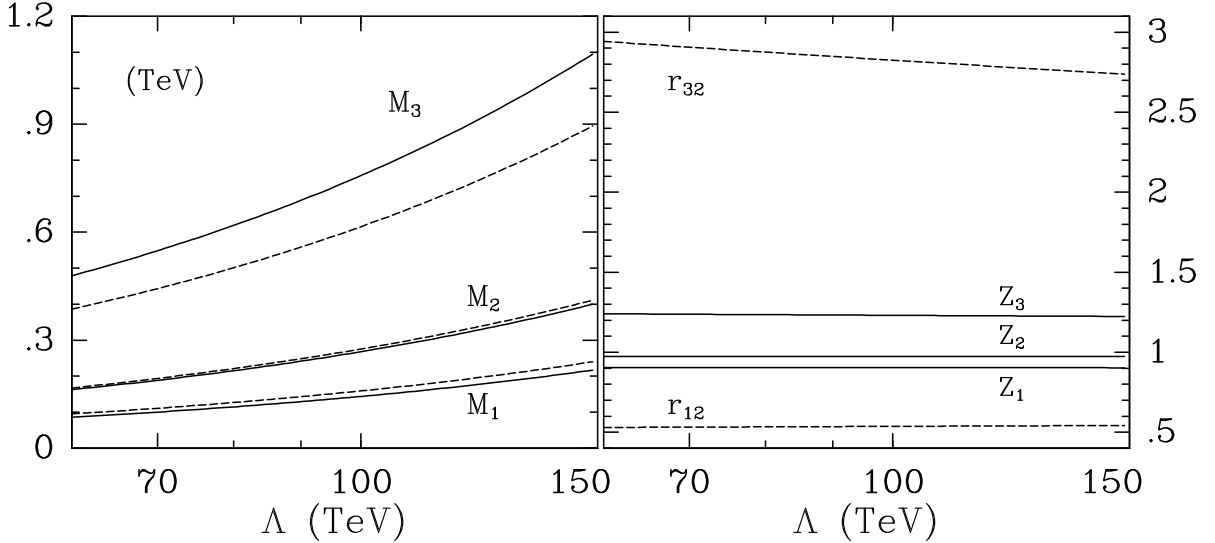}
\end{center}
\caption[f5]{\tenrm{Gaugino masses as a function of $\Lm$, solid
 lines. The dashed lines indicates the corresponding high--energy
 masses $M_i(X)$. On the right frame are shown the ratios 
 $r_{ij}$ and $Z_i$ defined in the text. 
}}
\label{gaugpar}
\end{figure}

It is a priori not obvious that the previous equations can yield 
physically meaningful results, due to the little freedom which the 
model allows and the smallness of the logarithms resummed by the 
RGE, i.e. 
of the period of integration $t_{XQ_0} = 2 \ln (X/Q_0)$. The peculiar 
interplay between strongly and weekly interacting sectors in this 
model, as we shall see, makes possible the breaking of the 
electroweak gauge group; maintains $B$ small at low--energy, therefore 
inducing large values of $\tanb$, and has consequences for 
the viability of the model itself.

The evolution of supersymmetric parameters for which yukawa couplings
do not play any role is independent of the breaking mechanism. 
Low--scale gaugino masses are: 
\beq
 {M_i} = {M_i(X)}\, Z_i 
       = \Lm \wal_i({Q_0}) \,g(x)\,;  \qquad 
  Z_i \equiv  \frac{\wal_i({Q_0})}{\wal_i(X)} = 
  (1 -\wal_i({Q_0}) \, b_i \, t_{XQ_0}) \,,
\label{gaug}
\eeq
where it is: 
$\wal_i({Q_0}) = \wal_i /(1 -\wal_i \, b^{SM}_i \, t_{{Q_0}Z}) $ and 
$t_{{Q_0}Z} \equiv 2 \ln{({Q_0} /M_Z)}$. 
(If no scale is specified, it is understood that the relevant variables
are low--energy variables ($\sim M_Z$), with evolution frozen at
$Q_0$ for the massive ones.)
For $\Lm = 100\,$TeV, the period of integration $t_{X{Q_0}}$ is 10.6 to 
be compared with the value $\sim 60$ in the MSSM for the same 
decoupling mass and the high--scale $X$ coinciding with 
$M_{GUT} \sim 3\times 10^{16}$. The values of the three gauge 
couplings at $X$ and $Q_0$ as obtained for $\Lm= 100\,$TeV, starting 
from our input values at $M_Z$, are:
\beq
\begin{array}{lll}
 \alpha_1(X)\ =0.0191\,; \quad & \alpha_2(X)\ =0.0331\,; 
                         \quad & \alpha_3(X)\ =0.0739\,, \\
 \alpha_1(Q_0)=0.0173\,; \quad & \alpha_2(Q_0)=0.0322\,; 
                         \quad & \alpha_3(Q_0)=0.0910 \,.
\end{array}
\eeq
The coefficients $Z_i$ are then: $0.90,0.97,1.23$ ($i=1,2,3$).

Gaugino masses, evolution coefficients $Z_i$ and ratios
$r_{ij} =  M_i/M_j = \alpha_i({Q_0})/\alpha_j({Q_0})$ are shown in 
Fig.~(\ref{gaugpar}) as function of $\Lm$. Although we analyze 
this model for all 
$\Lm$'s from $15$ to $150\,$TeV, we show in this figure only 
the region from $\Lm \sim 62\,$TeV, for reasons which will become
clear after the inclusion of one--loop corrections to $V_0$.
For lower $\Lm$'s, the values of $Z_i$ hardly deviate from those 
given above. Almost unchanged is also 
$r_{12}$, whereas $r_{32}$ reaches $3.3$ at $\Lm \sim 20\,$TeV. 
Again, the values of $Z_i$ are to be compared to those 
obtained in the MSSM for the same scale $Q_0$:
$ Z^{GU}_i = 0.44, 0.84,2.35 (i=1,2,3)$.
For each value of $\Lm$,
the gaugino sector in the MMM can be identified with that
in the MSSM for $M = \Lm \wal_i g(x)/Z^{GU}_i$.

The first two generations of scalar masses, whose boundary conditions
can be re--expressed in terms of low--scale gaugino 
masses (\ref{bcscal}) as: 
\beq
m_i^2(X)
 \ = \  2 \left\{ 
   C_3  \left(\frac{M_3}{Z_3}\right)^2   +
   C_2  \left(\frac{M_2}{Z_2}     \right)^2   + 
   \frac{3}{5} Y^2 \left(\frac{M_1}{Z_1}\right)^2   
         \right\}      \frac{f(x)}{(g(x))^2}\,, 
\label{bcm2}
\eeq
evolve according to (\ref{twogen}) in Appendix~A
and get corrected by quantities 
$\Delta {m_i}^2 \equiv {m_i}^2- {m_i^2(X)} $:
\beq
 \Delta m_i^2  =  
 \ 2 \left\{ 
  C_3    \left(   \frac{M_3}{Z_3} \right)^2            
  \!  \left(\! \frac{1\!-\!{Z_3}^2}{b_3}\!   \right) 
 +C_2    \left(   \frac{M_2}{Z_2} \right)^2            
    \left(\! \frac{1\!-\!{Z_2}^2}{b_2}\!  \right) 
 +\frac{3}{5} Y^2 \left( \frac{M_1}{Z_1} \right)^2 
  \!  \left(\! \frac{1\!-\!{Z_1}^2}{b_1}\!  \right)
             \right\} \,.            
\label{shiftsquark}
\eeq
For our choice of $x$, it is $f(x)/(g(x)^2) \sim 1 $. The 
corrections (\ref{shiftsquark}) amount to $\sim 15 \%$ in the 
squark sector and at most $5 \%$ in the case of sleptons, modestly 
enlarging the already wide gap existing between squark and slepton 
spectra. Dropping the indices of first and second generation, we 
have, for $\Lm = 100\,$TeV:
\beq   
   m_{\wti Q}^2 \simeq 16.8 M_2^2;              \quad  
   m_{\wti U}^2 \simeq m_{\wti D}^2 \simeq 15.4 M_2^2; \quad 
   m_{\wti L}^2 \simeq 1.64 M_2^2;              \quad 
   m_{\wti E}^2 \simeq 0.40 M_2^2 \,.
\label{coefMMM}
\eeq   
This large gap sharply distinguish the MMM 
from the MSSM, which in general predicts for the first two generations
of squarks and sleptons the familiar relations : 
$$
 m_{\wti Q}^2 \simeq m^2 \!+\! 6.4 M_2^2;\
 m_{\wti U}^2 \simeq m_{\wti D}^2 \simeq m^2 \!+\! 5.9 M_2^2; \
 m_{\wti L}^2 \simeq m^2 \!+\! 0.68 M_2^2; \
 m_{\wti E}^2 \simeq m^2 \!+\! 0.22 M_2^2. 
\label{coefMSSM}
$$
Therefore, whereas the gaugino sector in the two models can be
identified through a specific choice of $M$ for each value of $\Lm$, an
identification of the scalar mass parameters would require 
very different values of $m^2$ for squarks and sleptons 
($10.4 M_2^2$ and $0.96 M_2^2$ for the 
$SU(2)_L$ doublets; $9.5 M_2^2$ and $0.18 M_2^2$ for the 
singlets).

The interplay between such a light ``weak'' sector and the heavy 
``strong'' one is such to turn $m^2_{H_u}$ to large negative values 
as in the MSSM, making therefore possible the radiative breaking 
of $SU(2)_L\times U(1)_Y$. If, for the purpose of illustration,
we keep only the first of the logarithms to be resummed when 
solving (\ref{higgsrge}), we get:
\beq
 m_{H_u}^2 =  m_{\wti L}^2(X) - 3 \wal_t(X)
     \left(m_{\wti Q}^2(X) +m_{\wti U}^2(X)
                           +m_{\wti L}^2(X) \right) t_{X{Q_0}}
\label{humass}
\eeq
with $\alpha_t(X) = h^2_t(X)/(4 \pi)$ and $h_t(X)$ of 
${\cal O}(1)$. The factor $\sim 6$ lost in $t_{X{Q_0}}$ with respect 
to the MSSM is compensated in (\ref{humass}) by the heaviness of the 
squark spectrum. A similar effect si observed in the evolution of 
$m_{H_d}^2$. 

\begin{figure}[t]
\begin{center} 
\epsfxsize=12.0 cm
\leavevmode
\epsfbox[128 515 482 600]{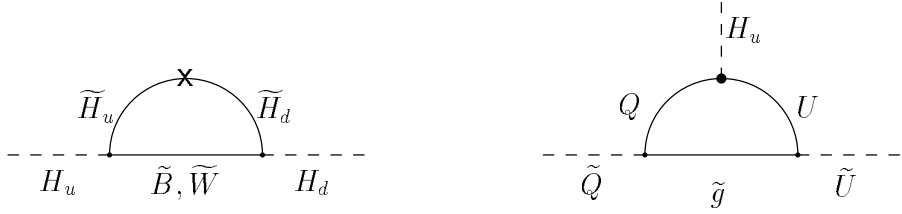}
\end{center}
\caption[f5]{\tenrm{One--loop diagrams contributing to 
$B \mu$ and $A_t$. Here and in the following figure, the cross 
indicates the parameter $\mu$, the dots one or two powers of 
yukawa couplings.}}
\label{diagone}
\end{figure}

Cancellations between weak and a higher--loop strong terms
appear also in the determination of the low--energy value of $B$.
Starting from the boundary condition
(\ref{bcaabb}), a value different from zero is generated 
at the one--loop level through gaugino mediated loops (see first 
diagram in Fig.~\ref{diagone}). The leading contributions are 
resummed in a series $\sum_n c^w_n (\wal(X) \, t_{X{Q_0}})^n$ where 
$\alpha(X)$ indicates generically $\alpha_2(X)$ or $\alpha_1(X)$.
In a closed form, it is:
\beq 
 B = - 4 \left\{
    C_2  \left( \frac{M_2}{Z_2} \right)          
    \left(\frac{1\!-\!{Z_2}}{b_2} \right) 
  + \frac{3}{5} \left(\frac{1}{4}\right)
            \left( \frac{M_1}{Z_1} \right)              
    \left(\frac{1\!-\!{Z_1}}{b_1} \right) 
                 \right\} \,. 
\label{Bgaug}
\eeq

Values of $A_i$ different from zero are similarly obtained from 
one--loop diagrams as the second one in Fig.~\ref{diagone}. 
Leading logarithms arising from such diagrams, after resummation, 
give: 
\beq 
 A_i = - 4 \left\{
    C_3   \left(\frac{M_3}{Z_3}\right)
    \left(\frac{1\!-\!{Z_3}}{b_3}\right) 
  + C_2   \left( \frac{M_2}{Z_2} \right)
    \left(\frac{1\!-\!{Z_2}}{b_2} \right) 
  +\frac{3}{5} \,C^Y_i 
           \left( \frac{M_1}{Z_1} \right)
    \left(\frac{1\!-\!{Z_1}}{b_1} \right)
                 \right\}\,.
\label{Agaug}
\eeq
The coefficients $C^Y_i$ are given in Appendix~A and $C_3$, $C_2$ are 
equal to $4/3,3/4$ for $A_t$ and $A_b$, and $0,3/4$ for 
$A_\tau$. When only gaugino--mediated loops are considered, it is 
then, $A_t \sim A_b \sim - M_2$ and $B \sim A_\tau \sim -0.1 M_2$, 
roughly for any $\Lm \gtap 70\,$TeV. 

\begin{figure}[t]
\begin{center} 
\epsfxsize=12.0 cm
\leavevmode
\epsfbox[128 420 482 620]{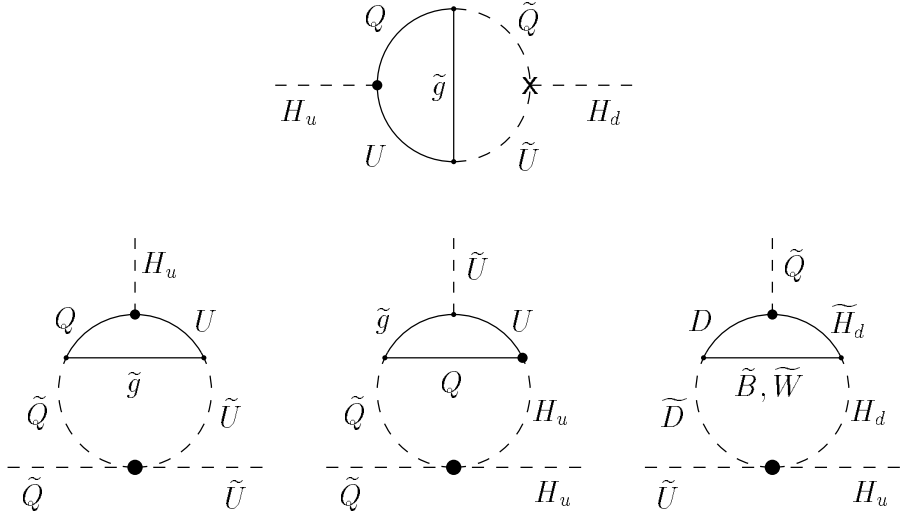}
\end{center}
\caption[f5]{\tenrm{Two--loop diagrams generating $B\mu$ and 
$A_t$. Both sets of diagrams contribute to the corresponding 
one--loop and two--loop RGE.}}
\label{diagtwo}
\end{figure}

Thus, the leading contributions to $B$ proportional to yukawa 
couplings, i.e. ``$A_i$--induced'', are first obtained at the 
two--loop level (see upper diagram in Fig.~\ref{diagtwo}) and give 
rise to the series
$\sum_n c^s_{n+1} \wal_t(X) (\wal_3(X))^n \, (t_{X{Q_0}})^{n+1}$. 
At the two--loop level are also obtained the leading logarithmic 
contributions to $A_i$ proportional to yukawa couplings (see lower 
diagrams in Fig.~\ref{diagtwo} for $A_t$). Nevertheless, while 
these diagrams produce indeed numerically small corrections to the 
values of $A_i$ previously obtained, the same is not true in the 
case of $B$, for which one has:
\beq
\frac{c_2^s}{c_1^w} \simeq
 - 3 \left(\frac{16}{9}\right) 
     \left(\frac{M_3}{M_2}\right)
     \left(\frac{Z_2}{Z_3}\right)
     \left(\frac{\wal_3(X)}{\wal_2(X)}\right) 
     \left(\wal_t(X) \, t_{X{Q_0}} \right); 
\quad \quad
\left \vert \frac{c_2^s}{c_1^w} \right \vert \gtap 1.  
\label{comp}
\eeq
Large cancellations between weak and higher--loop--order strong terms 
contributing to $B$ take place. They induce a flip of sign for this 
parameter, which, then, turns out to be small and positive. The 
positivity of $B$ forces also $\mu$ to be positive; its smallness, 
relative to the heavy scalar spectrum, pushes $\tanb$ to large values. 

At the electroweak minimum, after minimization of $V_0$, one 
obtains $\tanb = 46.4$ and the following values of yukawa 
couplings at the three relevant scales $X$, $Q_0$ and $M_Z$:
\beq
\begin{array}{lll}
 h_t(X) = 0.8130\,; \quad & h_t(Q_0) = 0.8755\,; 
                    \quad & h_t(M_Z) = 0.9799\,, \\ 
 h_b(X) = 0.6324\,; \quad & h_b(Q_0) = 0.7045\,; 
                    \quad & h_b(M_Z) = 0.7984\,, \\ 
h_\tau(X)=0.4812\,; \quad & h_\tau(Q_0)=0.4716\,;
                    \quad & h_\tau(M_Z)=0.4657\,.  
\end{array}
\eeq
The Higgs mass parameters $ m^2_{H_u}$ and $ m^2_{H_d}$ are:
$ m^2_{H_u} \simeq - (1.50 M_2)^2 \simeq -(0.53 M_3)^2$,  
$ m^2_{H_d} \simeq - (0.97 M_2)^2 $
and $\mu$, given practically by $-m^2_{H_u}$ 
(the difference between the two numbers determining $M_Z$), is 
$\mu \simeq 1.48 M_2$. The values of $A_t$ and $A_b$ do not deviate 
much from those obtained according to (\ref{Agaug}):
$A_t \sim A_b = -0.9 M_2 $. $A_\tau$ is now $-0.055 M_2$, and 
$B$ only $\sim 0.02 M_2$. 

\begin{figure}[t]
\begin{center} 
\epsfxsize=16.0 cm
\leavevmode
\epsfbox[75 507 540 720]{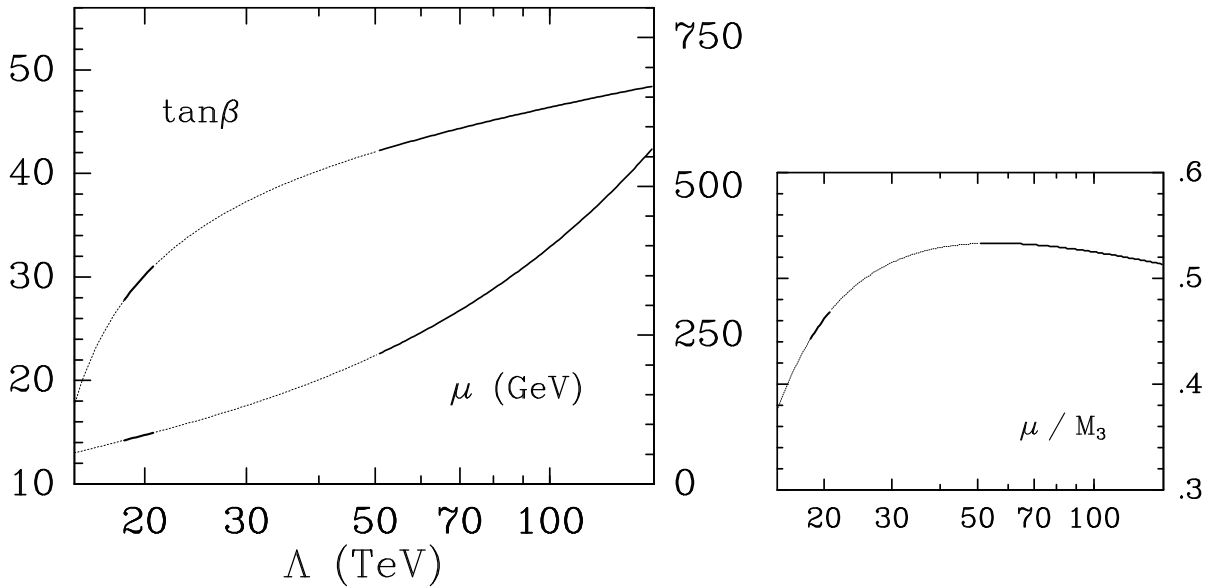}
\end{center}
\caption[f5]{\tenrm{Values of $\tanb$--$\mu$ obtained from the 
minimization of $V_0$. 
The dotted intervals indicate non--physical solutions.}}
\label{mmmfig}
\end{figure}

The results for $\tanb$--$\mu$ obtained from the minimization of 
$V_0$, as a function of $\Lm$ are shown in Fig.~\ref{mmmfig}. 
Some portions of these lines are dotted to indicate that 
the corresponding points of parameter space are non--physical. 
Negative squared masses are obtained in the spectra relative to 
these points, for scalar other than $H_u$ and $H_d$.  

The lightest low--energy soft parameter is certainly $m_{\wti E}^2$. 
For small values of $\Lm$, however,
the inclusion of $D$-terms renders sneutrinos lighter than charged
sleptons. In the leptonic sector, D--terms are: 
\beq
 D_{\widetilde \nu}   = \frac{1}{2}\cos 2 \beta {M_Z^2}\,; \qquad 
 D_{{\widetilde e}_L} = - D_{\widetilde \nu} 
                          ( 1 - 2\,\sin^2 \theta_W)\,; \qquad 
 D_{{\widetilde e}_R} = - D_{\widetilde \nu} 
                         \, 2\, \sin^2 \theta_W\,, 
\eeq
and numerically they amount to 
$ D_{\widetilde \nu}   \simeq - (64\,{\rm GeV})^2$, 
$ D_{{\widetilde e}_L} \simeq  
  D_{{\widetilde e}_R} \simeq (45\,{\rm GeV})^2$ 
for all values of $\tanb$ obtained here through minimization of 
(\ref{higgspot}). 
Moreover, if $\Lm$ is small enough, $m_{{\wti \nu}_i}^2$ may become
negative. By using (\ref{gaug})--(\ref{coefMMM}) (i.e. neglecting, 
for a rough estimate, third generation mass effects on the evolution 
of soft parameters) and our gauge couplings inputs, it is easy to see 
that the condition 
$m_L^2 > | D_{\widetilde \nu} |$,
i.e. $m^2_{\widetilde \nu}> 0$, 
is fulfilled for 
$ \Lm > 7.3 \,|\cos 2 \beta |^{1/2} M_Z/\alpha_2 \simeq 20\,$TeV. This 
value is not too dissimilar from that obtained with an exact 
calculation, i.e. the value of $\Lm$ in Fig.~\ref{mmmfig} where the 
initial dotted intervals of the two curves for $\mu$ and $\tanb$ 
turn into solid ones.

In this same lowest range of $\Lm$, where it is $m_{{\wti \nu}_i}^2<0$, 
$D_{{\widetilde e}_L}$ and $D_{{\widetilde e}_R}$ protect the squared 
masses of charged sleptons from becoming negative. Nevertheless, the 
mass of the lightest state, ${\wti \tau}_1$ is smaller than the 
experimental lower bound of $45\,$GeV coming from LEP\,I and it remains 
so also in the tiny region of $\Lm$ around $20\,$TeV, where $\tanb$ 
and $\mu$ are indicated by solid lines.

When $\Lm$ increases, sneutrinos become heavier, whereas 
$m_{{\wti \tau}_1}$ tends initially to decrease. This is due to the 
fast increase of $\mu \, \tanb$ 
in the left--right entry of the ${\wti \tau}$ mass matrix,
initially faster than the increase of the diagonal entries (see
the $\Lm$ dependence of $\mu/M_3$ in Fig.~\ref{mmmfig}). By 
identifying $D_{{\widetilde e}_L}$ and $D_{{\widetilde e}_R}$
($ \equiv D_{{\widetilde e}}$), and neglecting third generation 
mass effects on the evolution of soft parameters, as well as 
$A_\tau$ in the off--diagonal entry of the 
${\wti \tau}$ mass matrix, one can cast the condition 
$m^2_{\wti \tau_1} \le 0$ in the form:
$  m^2_E \, m^2_L + 
 D_{{\widetilde e}} (m^2_E + m^2_L )
 - ( \mu \,\tanb m_\tau )^2  +  D_{{\widetilde e}}^2 
 \le 0 $. 
This quadratic equation in $M^2_2$ admits indeed two solutions 
which delimit 
an interval not too dissimilar from that indicated by the dotted 
portions of lines in Fig.~\ref{mmmfig} and determined without 
any of the above approximations. For further increases in $\Lm$, 
the diagonal entries in the ${\wti \tau}$ mass matrix 
increase far more rapidly than the off--diagonal
ones giving eventually a physically acceptable spectrum. 

We observe that choices of $x$ closer to 1 (i.e. $X \sim \Lm $), 
affect the supersymmetric spectrum through the 
functions $f(x)$, $g(x)$ and the smaller size of the logarithm 
$t_{XQ_0}$. We obtain lighter squarks and sleptons as well as 
lighter values of $|m_{H_u}^2|$: $\mu$ is, then, in general, 
$\sim M_2$. All gaugino are heavier and therefore weak and 
strong contributions to $B$ are also larger. In the range of 
$\Lm$ physically acceptable ($\Lm \gtap 50\,$GeV), the values of $B$ 
are smaller than those obtained with our previous choice of $x$ (see 
(\ref{comp})): strong gauge couplings as well as top yukawa couplings 
are now larger at $X$, but not enough to compensate the decrease of  
$t_{XQ_0}$ due to smaller $X$'s for fixed $\Lm$'s. (The values of
$Q_0$ are also slightly smaller, but this change has negligible 
consequences.) Overall, the new $\tanb$'s do not differ appreciably 
from the values previously obtained, for $\Lm \gtap 50\,$GeV.  

\subsection{Fully one--loop corrected Higgs potential}

We have previously argued about the suitability of our choice of 
decoupling scale $Q_0$, which will be a posteriori justified by the 
results of the calculations presented in this subsection.
It is obvious, however, that a change in this scale affects those 
parameters where the interplay between weak and strong sectors with 
different sensitivity to $Q_0$ has the largest effect. Particularly
problematic is the parameter $B$. Since its low--energy value turns
out to be rather small, changes of $Q_0$ may easily induce 
oscillations of $B$ around zero, which in turn, require flips of 
sign in the parameter $\mu$ with non--negligible consequences 
for the resulting phenomenology. Higher order corrections than those
provided by the one--loop RGE as well as adjustments for the  
inadequacy of one unique decoupling scale $Q_0$ for the widely 
separated strong and weak sectors, become therefore mandatory. 

Given the smallness of the logarithm $t_{{Q_0}X}$, finite 
corrections are in the MMM more important than in the MSSM, when 
compared to the leading logarithmic corrections supplied by the 
same order RGE. In principle, they should be also larger than 
corrections coming from higher order RGE. In practice, 
this will not be the case, for the parameter $B$. 

For a complete set of corrections to the parameters which determine 
the radiative breaking of $SU(2)\times U(1)$ we will consider the 
fully one--loop corrected effective potential $V_1(Q_0)$, defined as 
$V_1(Q_0) \equiv V_0(Q_0)+\Delta V_1(Q_0)$ and we will include 
contributions to $\Delta V_1(Q_0)$ coming from all sectors of the 
theory. Because of the wide gap existing between weak and strong mass 
parameters, this is crucial in order to warranty results stable 
under variation of the decoupling scale. 

Among other possible finite corrections, we include those to the
b--quark mass, which, as known, can be rather large, when $\tanb$ has 
values as large as those obtained here. These corrections reduce the 
value of the coupling $h_b$ since 
$h_b(M_Z) = (m_b - \delta m_b )/ v_d = 
            (m_b - \tanb \mu \Delta )/ (v \cos \beta )$ 
with $\Delta $ ($> 0$) given in~\cite{CAPOWA,RASA}. Consequently, they 
affect the evolution of $m_{H_d}^2$ (see (\ref{higgsrge})), 
decrease the low--energy value of $B$, therefore increasing 
$\tanb$. This, in turn, increases $1/\cos\beta $ and the size of 
$\delta m_b$. A retuning of all parameters finally produces the 
electroweak minimum for $\tanb$'s in general larger than those 
obtained when these corrections were not included. The values of 
$\mu$ remain practically unchanged, since the parameters $m_{H_u}^2$ 
are unaffected by these corrections. For $\Lm = 100\,$TeV, 
in particular, 
it is $\mu \sim 397.5\,$GeV, as before; $h_b$ decreases from 
0.798 to $0.786$ and correspondingly $A_t$ and $A_b$ are slightly 
more negative. $B$ adjusts to values indistinguishable from the 
previous one, but $\tanb$ increases to $52.1$ from the initial 
value of $46.4$. We neglect finite corrections to quark masses 
other than to $m_b$, 
as well as threshold effects for supersymmetric parameters other 
than those induced by $\Delta V_1(Q_0)$. 

We come now to analyze the effects of the inclusion of 
the one--loop corrections to the effective potential, 
which have the well known expression:
\beq
 \Delta  V_1(Q_0)  =    \frac{1}{64 \pi^2}   \sum_{a} 
  n_a m^4_a(H) \, \left[ \ln \left(\frac{m^2_a(H)}{Q_0^2}\right) 
                                 - \frac{3}{2} \right]
\label{higgscorr}
\eeq
where $m_a(H)$ is the 
field-dependent mass of the $a^{th}$-particle, $n_a$, the 
corresponding number of degrees of freedom (negative for 
fermions): $ n_{\wti q} =  6$, $ n_{q} =-12$, $ n_{\wti l} =2$, 
$n_l = -4$, $n_{{\wti \chi}^+} = - 4$, $ n_{H^\pm} = 2$, $n_W = 6$, 
$n_{{\wti \chi}^0} = - 2$, $ n_{H^0} = 1$, and $n_Z = 3$. 
The corrected potential $V_1(Q_0)$
yields minimization equations which retain the form in 
(\ref{derivtreeone}), (\ref{derivtreetwo}), provided the parameters 
$B\mu$, ${m}^2_{H_d}$, ${m}^2_{H_u}$ are substituted by 
hatted parameters ${\widehat {B\mu}}$, ${\widehat m}^2_{H_d}$, 
${\widehat m}^2_{H_u}$ defined as 
${\widehat m}^2_{H_i} = m^2_{H_i} + \delta m^2_{H_i}$, and 
${\widehat{ B \mu}} = B \mu + \delta (B\mu)$:
\bea
 \left( 
  {\widehat m}^2_{H_d} + \mu^2 + \nhalf\, \ggg ({v_d^2-v_u^2})
 \right) 2 v_d  - ({\widehat {B\mu}})\, 2 v_u      & = & 0   
\label{derivcompo}                                 \\
 \left( 
  {\widehat m}^2_{H_u} + \mu^2 - \nhalf\, \ggg ({v_d^2-v_u^2}) 
 \right) 2 v_u  - ({\widehat {B\mu}})\, 2 v_d      & = & 0\,. 
\label{derivcompt}
\eea
The shifts $ \delta m^2_{H_i}$, $ \delta (B\mu)$ are listed in 
Appendix~B. We do not use different symbols to distinguish between $v_u$ 
and $v_d$ as obtained from the minimization of $V_0$ and $V_1$. In 
what follows and in Appendix~B we refer to the new minimum as to the 
$V_1$--minimum as opposite to the $V_0$--minimum obtained from 
(\ref{higgspot}).

The corrections $\delta m^2_{H_i}$ have been only partially included
in~\cite{DIM,BAGG}, but not those to $B \mu$, in principle very 
important and with strong impact on the value of $\tanb$. In a 
generic GMSB, one can 
fix the value of $\tanb$ at will using the freedom in the high--
and low--energy parameter $B$. It is precisely this lack of 
freedom which makes the MMM a model far more difficult to study. 

At the $V_1$--minimum, quark/squark and lepton/slepton contributions 
to $\delta (B\mu)$ 
are~\footnote{Although squark mass matrices were evolved down in 
their $6\times 6$ form, the approximation of $2\times 2$ mass 
matrices for each generation of squarks and sleptons is used here}: 
\beq
  \frac{1}{32\pi^2} \left[ 
 3  \left(A_t\, \mu \, h_t^2 \right) D(m_{{\wti t_1},{\wti t_2}}) +
 3  \left(A_b\, \mu \, h_b^2 \right) D(m_{{\wti b_1},{\wti b_2}}) +
    \left(A_\tau\, \mu \, h_\tau^2 \right) 
                                     D(m_{{\wti \tau_1},{\wti \tau_2}}) 
                    \right]\,; 
\label{Bscal}
\eeq 
those coming from charged gauge/gaugino, higgs/higgsino modes:
\beq
  \frac{1}{32\pi^2} \left[
  2 \left(M_2\mu \, g^2 \right) D(m_{{\wti \chi_1},{\wti \chi_2}})
  \,  - \frac{g^2}{4} 
  \left(2 (B\mu) + \frac{g^2}{2} v^2 \sin 2 \beta  \right)
                     \, D(m_{H^\pm,G^\pm})
                    \right]\,, 
\label{Bcharg}
\eeq
where $v^2 \equiv v_d^2 + v_u^2 $ and $G^\pm$ is what would be the 
charged Goldstone boson at the $V_0$--minimum. Those due to 
neutral gauge/gaugino, higgs/higgsino mode are: 
\beq
  \frac{1}{32\pi^2} \left[
  2 \,\sum_{i=1}^4  F(m_{{\wti \chi_i}^0},Q_0) 
    \, \frac{\partial m^2_{{\wti \chi_i}^0}}{\partial \sin 2 \beta}   
  - \frac{\ggg}{4}  
      \left( (B\mu) + \frac{\ggg}{4} v^2 \sin 2 \beta \right)
                            \, D(m_{A^0,{G^0}}) 
                    \right] \,,
\label{Bneutr}
\eeq
with $A^0$ the pseudoscalar Higgs, 
$G^0$ the would--be--neutral Goldstone boson. The functions 
$D(m_{a,b})$ and $F(m_a,Q_0)$ are defined in Appendix B. Corrections to 
$B\mu$ coming from other sectors of the theory are identically zero. 

A comparison of (\ref{Bscal}) with the RGE for $B$ (see~(\ref{brge})) 
shows how the quark/squark, lepton/slepton contributions to 
$\delta (B\mu)$ are corrections to the parameter $B$. They improve
upon the arbitrariness of the scale $Q_0$, linking it more 
realistically to the actual mass of the scalars virtually exchanged 
in the corresponding loop diagram. In the top/stop case, for example, 
in the limit of vanishing left--right mixing terms, it is indeed, 
$m_{\wti t_1}  \to m_{\wti U}$ and $m_{\wti t_2}  \to m_{\wti Q}$, 
with ${\wti U}$, ${\wti Q}$ the two scalars in the first diagram of 
Fig.~3. Moreover, in the limit 
$m_{\wti t_1}  \to m_{\wti t_2} \equiv m_{\wti t}$, it is 
$D(m_{\wti t_{1,2}} ) \to 2 \log (m^2_{\wti t}/Q_0^2)$ and the term 
$ 3/(16 \pi^2) A_t h_t^2 \log(m^2_{\wti t}/Q_0^2)$, with opposite 
sign to the corresponding one arising from (\ref{brge}), would have 
the effect of trading $Q_0^2 $ for $m^2_{\wti t}$. In its actual form, 
the top/stop contribution includes also finite corrections, of type 
$ c'^s_2 \,\wal_t(X) \,\wal_3(X) \, t_{XQ_0}$ i.e. with one logarithm
less than the corresponding term provided by the RGE
$c^s_2 \,\wal_t(X) \,\wal_3(X) \, (t_{XQ_0})^2$. 

Similar considerations hold for chargino and neutralino corrections 
which take care of scale adjustments and finite pieces inclusions, 
of type $\sim c_1'^{\,w} \,\wal(X) $, coming from the first diagram of 
Fig.~\ref{diagone}, to be compared to the terms 
$ c_1^w \,\wal(X) \, t_{X{Q_0}}$ obtained from the one--loop RGE. 

The sign of the contributions (\ref{Bscal})--(\ref{Bneutr})
depends crucially on the value of $Q_0$ with respect to the masses 
exchanged in the loop. Keeping the choice of $Q_0$ made at the 
tree--level, we have for $\Lm = 100\,$TeV, 
$ m_{\wti t_1} \sim 977\,$GeV, 
$ m_{\wti t_2} \sim 1099\,$GeV, and $Q_0 \sim 993\,$GeV. These 
masses, roughly independent of further adjustments in $\tanb$ and
$\mu$, are shown in Fig.~\ref{spectrum} normalized to the gluino 
mass. 
Thus, $D(m_{\wti t_{1,2}})$ is positive; the same is true for 
$D(m_{\wti b_{1,2}})$, whereas $D(m_{\wti \tau_{1,2}})$ and 
$D(m_{\wti \chi_{1,2}})$ are negative. Quark/squark corrections 
are therefore small and negative ($A_t$ and $A_b$ are both negative), 
those due to tau/stau are small and positive. Chargino corrections 
are large and negative. Although this cannot be explicitly seen by 
the formulas displayed in (\ref{Bneutr}) and in Appendix B, negative are 
also the neutralino contributions to $\delta (B\mu)$, although not as 
large as the chargino contributions. The remaining corrections coming 
from gauge and higgs boson modes are numerically smaller. 

Whereas the overall $Q_0$--dependence of ${\widehat {B\mu}}$ and 
$\delta m_{H_i}^2$ is of order higher than the order of the 
calculation presented here, the relative size of the individual 
corrections changes for different choices of $Q_0$. For smaller 
values of $Q_0$, the quark/squark contribution tend to dominate and 
it may happen that approaching $M_Z$, a negative $\mu$ is 
needed to balance corrections too negative and therefore maintain 
${\widehat {B\mu}}$ positive (${\widehat {B\mu}}$ is indeed related to 
the fully one--loop corrected pseudoscalar Higgs mass). Since the bulk 
of supersymmetric particles is much heavier than $M_Z$, we believe that
definite conclusions cannot be drawn on the solutions obtained with 
such an ``unnatural'' choice of scale without including higher order 
corrections to the minimization conditions. 

\begin{figure}[t]
\begin{center} 
\epsfxsize=16.0 cm
\leavevmode
\epsfbox[75 507 540 720]{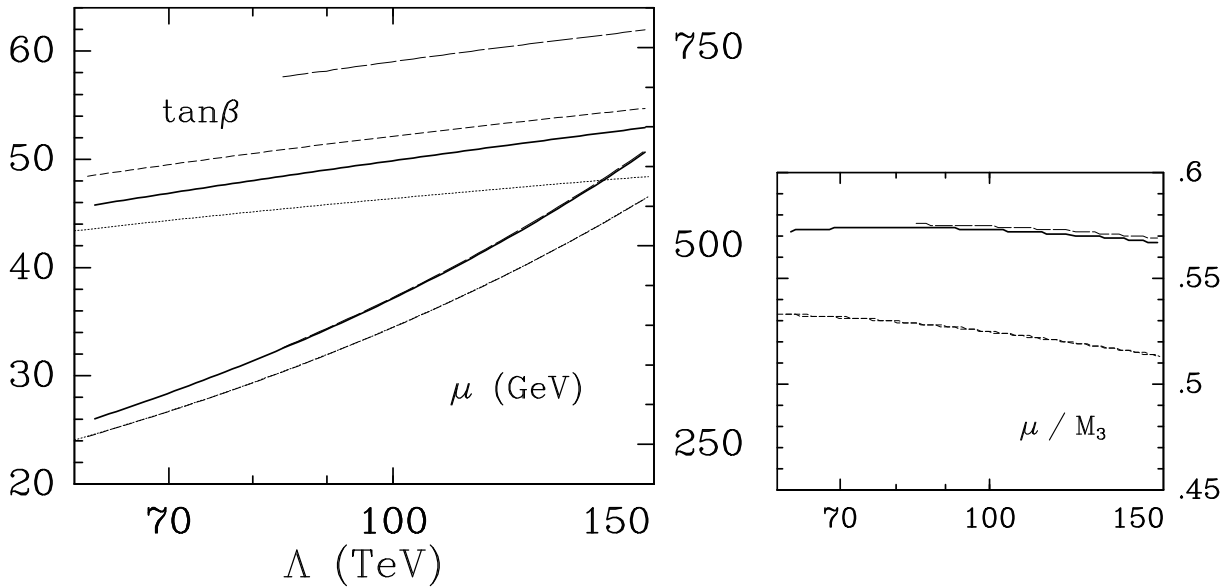}
\end{center}
\caption[f5]{\tenrm{Solutions obtained for $\tanb$--$\mu$ 
with a fully one--loop corrected neutral higgs potential (solid lines).
The short--dashed lines correspond to the solutions obtained 
when only $m_b$--corrections are included; the dotted lines
are the solutions in Fig.~\ref{mmmfig}. See text for the 
long--dashed lines. 
}}
\label{mmmfigloop}
\end{figure}

As for the shifts $\delta m_{H_i}^2$, the two largest contributions 
are the top/stop correction to $m_{H_u}^2$ and the bottom/sbottom 
correction to $m_{H_d}^2$. They are both negative and add to the 
already negative parameters $m_{H_i}^2$. Contributions from first 
two generations squarks with different isospin, although not 
negligible in absolute value, almost completely cancel each other 
(see formulas in Appendix~B). The next largest contributions are the 
chargino/charged--boson--sector and neutralino/neutral--boson 
sector, which contribute with different sign. Their algebraic sum, 
positive, is however still larger than the tau/stau 
contribution.  
The hatted parameters ${\widehat m}^2_{H_d}$, ${\widehat m}^2_{H_u}$ 
are now more negative than $m^2_{H_d}$ and $m^2_{H_u}$.

A retuning of $\mu $ and $\tanb $ is therefore needed to make the 
left--hand--side of (\ref{derivcompo}) and (\ref{derivcompt})
vanish. After a numerical iteration, solutions are found for 
values of $\tanb$ and $\mu$ larger than those obtained at the 
$V_0$--minimum. Considerable is, in particular, the deviation 
for $\tanb$: for $\Lm =100\,$TeV, the new value is $59.0$, the old,
$46.4$. This situation is not improved by a change of 
decoupling scale, which is symptomatic of some 
incompleteness in the set of corrections included. 

A close inspection of the two--loop evolution equation for $B$ (see 
the relative equation in Appendix~A) shows the presence of strong 
terms, not $A_i$--induced, which yield contributions of type 
$ d_{o\,n}^s \, (\wal_s(X) \,\wal_t(X) \, t_{XQ_0} )^n $. The first 
of these terms, with $n=1$, is of same size of the 
$A_i$--induced finite corrections. It is indeed originated by the 
same upper diagram in Fig.~\ref{diagtwo}, with the fermionic loop 
``open'' (i.e. not shrank to reproduce $A_t$). 
We add this first 
term in our determination of $B$; it has a positive sign and produces
therefore an increase in the value of $B$ and a decrease of 
$\tanb$. 

In complete analogy with the situation observed in the case of the 
one--loop RGE, 
the remaining strong $A_i$--induced terms in the two--loop RGE 
yield corrections of type
$ d_{1\,n}^s \, (\wal_s^2(X) \,\wal_t(X) \, t_{XQ_0}^2 )^n $ and 
$ d_{2\,n}^s \, (\wal_s(X) \,\wal_t^2(X) \, t_{XQ_0}^2 )^n $, 
of same size than the corrections 
$ d^w_n (\wal^2(X) t_{XQ_0})^n$ due to weak terms. 
The first terms in these series ($n=1$), are larger than the 
corresponding finite one--loop corrections,
$c^{'\,s}_1 \,\wal_s(X) \,\wal_t(X) t_{X{Q_0}}$, 
$c^{'\,w}_1 \, \wal(X)$, as expected. 
Their addition, therefore, would certainly 
require the inclusion of finite thresholds effects and corrections
due to the two--loop RGE for all supersymmetric parameters.
No other anomalously large corrections are induced by terms in the 
two--loop RGE for $m_{H_u}^2$ and $m_{H_d}^2$~\cite{MAVAU}.

We show in Fig.~\ref{mmmfigloop} (solid lines) our estimate of $\mu$, 
$\tanb$ as obtained following the prescription described above. We 
start our plots from 
$\Lm > 62\,$TeV~\footnote{This value of $\Lm$ is larger than 
the value of $\sim 50\,$TeV obtained when minimizing $V_0$, since
$\mu$ and $\tanb$ are now larger}; below this value it is 
$m_{\wti \tau_1}^2 <0 $, and 
the calculation of one--loop corrections becomes impossible (see terms 
$\ln(m_{\wti \tau}^2/Q_0^2)$ in the tau/stau corrections). We 
also neglect a very tiny region at small $\Lm$ which yields 
unacceptably low masses for ${\wti \tau}_1$. For comparison, we give in 
this figure also the results obtained through minimization of 
$V_0$ (dotted lines); those obtained when only finite corrections 
to $m_b$ are added (short--dashed lines) and those coming from 
the minimization of $V_1$, with finite corrections to $m_b$ included
(long--dashed lines). In this last case $\tanb$ is large enough to 
keep $m_{\wti \tau_1}^2$ negative up to $\Lm \sim 80\,$TeV. As 
already mentioned, corrections to $m_b$ only do not produce deviations
in the value of $\mu$ as obtained at the $V_0$--minimum (the 
dotted and short--dashed lines corresponding to $\mu$ coincide).
Similarly the inclusion of the non--resummed term in the 
two--loop RGE lowers $\tanb$, but gives rise to changes in $\mu$
indistinguishable on the left frame, and only barely visible in the 
right one, where the ratios $\mu/M_3$ are given. 

The addition of all the above described corrections to the Higgs
potential produce a much milder dependence of the $\tanb$--$\mu$ 
solutions on the decoupling scale $Q_0$. We 
give in the table below the values of $\tanb$ and $\mu$ at the 
$V_0$-- and the $V_1$--minimum for $\Lm = 100\,$TeV, and 
the two different 
decoupling scales $Q_0=993\,$GeV and $Q_0=331\,$GeV.

\begin{center}
\begin{tabular}{r|c|c}
                 & \qquad $Q_0 =993\,$GeV \qquad 
                 & \qquad $Q_0 =331\,$GeV \qquad 
   \\ \hline 
$V_0$--min \quad & \quad $\tanb =46.4$,\quad $\mu =397.5\,$GeV \qquad  
                 & \quad $\tanb =36.1$,\quad $\mu =481.8\,$GeV \qquad
   \\ \hline
$V_1$--min \quad & \quad $\tanb =49.9$,\quad $\mu =433.9\,$GeV \qquad  
                 & \quad $\tanb =47.3$,\quad $\mu =446.0\,$GeV \qquad  
   \\ \hline
\end{tabular}
\end{center}

\noindent 
The large variation obtained when minimizing $V_0$ can be traced 
back rather easily. The lowering of $Q_0$ induces an increase of the 
value of the strong gauge 
coupling in the interval \{$Q_0,X$\}. This is due to the fact that 
the interval \{$M_Z,Q_0$\} where the faster SM evolution takes 
place ($b^{SM}_3=-7$; $b_3 = -3$) is now smaller. The trilinear 
couplings $A_t$, $A_b$ are larger, and, as a consequence, also the 
values of $B$ increase. 
Heavier squark spectra and larger top yukawa couplings induce also 
larger $\mu$'s; the increase of the term $B\mu$ produces smaller 
values of $\tanb$. The variation in the $\tanb$--$\mu$ solution 
is much smaller when corrections to $V_0$ are included. 
Values of $Q_0$ too close to $M_Z$, are however still 
problematic.

This shows that our choice of $Q_0$ is already close enough to 
the optimized scale $Q_S$ discussed in~\cite{GRZ,CASAS}, where
the prediction for the two vacuum expectation values $v_1$ and
$v_2$ from $V_0$ and $V_1$ coincide. We can therefore rely on the 
RGE--improved tree--level scalar potential and our choice of 
$Q_0$ to verify whether dangerous minima breaking charge and/or
colour are not also present and possibly deeper than the observed
electroweak minimum. An 
analysis of this type when all one--loop corrections to the 
scalar potential are added would clearly be prohibitive. We do not 
find evidence for the existence of such minima, at least using
the analytical criteria given in~\cite{CASAS,RASA}.

\section{Spectrum, Prediction, Constraints}

\begin{figure}[p]
\begin{center} 
\epsfxsize=14.5 cm
\leavevmode
\epsfbox[75 75 555 715]{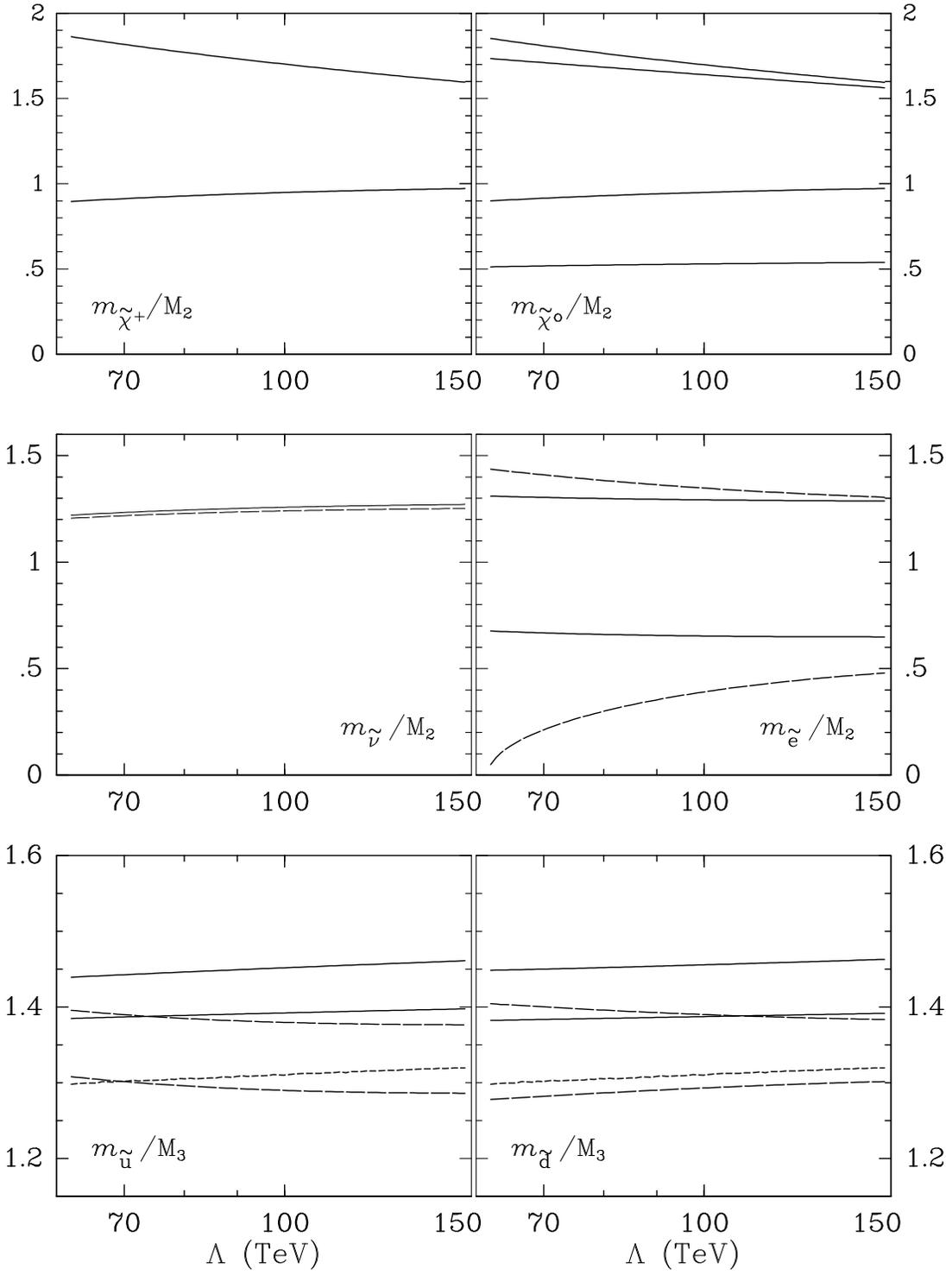}
\end{center}
\caption[f5]{\tenrm{Ratios of chargino, neutralino, and sfermion
 masses over weak and strong gaugino masses. For sleptons and
 squarks, the solid lines are relative to the degenerate first two
 generations left--handed (upper lines) and right--handed (lower
 lines) states. The dashed lines refer to the third generation
 eigenstates. The short--dashed lines in the two lower frames
 indicate the ratio of our choice for $Q_0$ (see text)
 over the gluino mass. This spectrum corresponds to the solutions
 ($\tanb,\mu$) indicated in Fig.~\ref{mmmfigloop} by solid lines.}}
\label{spectrum}
\end{figure}

We display in Fig.~\ref{spectrum} the spectrum obtained in the 
allowed range of $\Lm$ for our estimate of $\mu$, $\tanb$ (solid 
lines of Fig.~\ref{mmmfigloop}), and we give explicitly in the 
following table the values of gaugino, 
chargino. neutralino, and slepton masses (in GeV) at 
$\Lm=100\,$TeV.

\noindent 
\begin{tabular}{ccccccccccccccc}
 $M_1$                             & 
 $M_2$                             & 
 $M_3$                             & 
 $m_{{\wti \chi}^+_2}$             &
 $m_{{\wti \chi}^+_1}$             &
 $m_{{\wti \chi}^0_1}$             & 
 $m_{{\wti \chi}^0_2}$             &
 $m_{{\wti \chi}^0_3}$             &
 $m_{{\wti \chi}^0_4}$             &
 $m_{{\wti \nu}_{1,2}}$            &
 $m_{{\wti \nu}_3}$                &  
 $m_{{\wti \tau}_1}$               &   
 $m_{{\wti \tau}_2}$               &
 $m_{{{\wti e}_R}}$ &
 $m_{{{\wti e}_L}}$  
 \\[1.001ex] \hline   \\[-2.0ex]
 144 & 268 & 757 & 254 & 456 & 142 & 255 & 440 & 455 &
 337 & 333 & 105 & 361 & 175 & 347 
 \\[0.90ex] \hline 
\end{tabular} 

\noindent 
We follow here the conventions in~\cite{HAKA} 
($m_{{\wti \chi_1}^+} > m_{{\wti \chi_2}^+}$) and we indicate
by $m_{\wti e_R}$ ($m_{\wti e_L}$) the two degenerate eigenvalues
for right-- and left--handed ${\wti e}$ and ${\wti \mu}$. Similarly
degenerate are two of the three ${\wti \nu}$ states. For the same 
$\Lm$, the up-- and down--squark masses are given in Appendix~C. 
Due to the rather large values of $\mu$ and $\tanb$, the three 
heavy Higgs states 
(two neutral, one charged) range between $346$ and $360\,$GeV;
the lightest neutral state has mass roughly $125\,$GeV.

The almost factor of two between the 
first two generations left--handed and right--handed masses, already 
present at high--scale (\ref{bcm2}) is clearly visible in the fourth 
frame of Fig.~\ref{spectrum}. 
The same splitting in the masses of the first two generations squarks 
is more modest since it is due to weak gaugino loops, whereas the 
bulk of the masses is produced by gluino loops~(\ref{bcm2}). The 
corresponding ratios of the high--scale first two generations 
left-- and right--handed squark masses ($m_{\wti Q}(X)$, 
$m_{\wti U}(X)$) over the 
low--energy gluino mass, not shown in this figure, are shifted 
downward, with respect to the solid lines in the two lower frames, 
by a factor $\sim (1-Z_2^2)/b_3$ (for $f(x)/g(x)^2 \sim 1$) 
(see~(\ref{shiftsquark})). Their geometrical mean, $Q_0/M_3$, is 
explicitly indicated in both frames by the short--dashed lines. Tiny 
differences in the shape of the two solid lines in these two frames
are due to isospin effects (\ref{bcm2},\,\ref{shiftsquark}) as well 
as to the presence of different D--terms in the two cases. Those 
induced by intergenerational as well as chirality mixing terms in the 
up-- and down--squark mass matrices are not visible in these figures. 

Right--handed as well as left--handed ${\wti e}$ and ${\wti \mu}$ can 
easily decay to the lightest neutralino ${\wti \chi_1}^0$ with mass 
$\sim M_1$. The decrease for increasing $\Lm$ of the ratios relative
to the two heavy neutralinos (mainly neutral higgsino states) and 
the heavy chargino (mainly a charged higgsino state) in the 
first two frames of Fig.~\ref{spectrum}, can be simply 
explained by the milder growth of $\mu$ already observed in 
Fig.~\ref{mmmfigloop}. Their masses actually 
increase and their states become increasingly more mixed with the 
gaugino states. 

Third generation yukawa effects in the evolution equations 
(\ref{thirdgen}) are responsible for the roughly $10\%$ decrease in 
the third generation squark masses. In addition, the two stop masses 
are only a little affected by the presence of left--right mixing terms 
in the up--squark mass matrix ($\sim A_t m_t \ltap M_2^2 $ compared to 
the diagonal elements which are $\gtap M_3^2$). This effect is 
somewhat larger in the down--squark case. Differently than in the MSSM, 
the twelve squark states cluster quite closely around a common value of 
mass, $\sim 1 \,$TeV at $\Lambda =100\,$TeV. We show in Appendix~C the 
diagonalization matrices for up-- and down--squarks, at this same value 
of $\Lm$, from where one can read off the composition of the 
relative squark mass eigenstates.

Small is the effect induced by $h_\tau$ in the sneutrino sector, as the 
third frame of Fig.~\ref{spectrum} shows. On the contrary, the effects 
due to the presence of left--right mixing terms in the charged slepton 
mass matrix are rather large. The two states ${\wti \tau_1}$, 
${\wti \tau_2}$, have masses on opposite sides of the left-- and 
right--handed first two generations sleptons. The lighter of the two, 
${\wti \tau}_1$, is indeed the lightest sparticle in the spectrum and 
exceeds the LEP~I bound of $45\,$GeV only for $\Lambda \gtap 72\,$TeV. 
Once this constraint is imposed, the remaining spectrum is heavy 
enough to largely exceed any other experimental lower bound on 
masses. We obtain, in fact, 
$m_{{\wti \chi}^\pm}\gtap 180\,$GeV,
$m_{{\wti \chi}^0}  \gtap 100\,$GeV,
$m_{{\wti \nu}}     \gtap 240\,$GeV,
$m_{{\wti l}^\pm}   \gtap 130\,$GeV,
$m_{{\wti u}}       \gtap 735\,$GeV, and 
$m_{{\wti d}}       \gtap 725\,$GeV.

Thus, the main decay for ${\wti \chi_1}^0$ is the two body decay 
${\wti \chi_1}^0 \to \tau {\wti \tau}_1$. It proceeds with full 
electroweak gauge strength reducing therefore the partial width for 
the decay ${\wti \chi_1}^0 \to \gamma {\wti G}$ (${\wti G}$ is here 
the gravitino) to practically negligible levels in this model. 
(For a theoretical discussion of this decay mode 
see~\cite{CDFPHPHTH}.) The subsequent decay 
${\wti \tau}_1 \to \tau {\wti G}$ gives $\tau \tau$ +missing energy 
as signature for the decaying ${\wti \chi_1}^0 $ and a four $\tau$'s 
signal for ${\wti \chi}_1^0$ pair production in an $e^+e^-$ collider.
Given the rather large values of $m_{{\wti \chi}^0_1}$ induced by 
the constraint $m_{{\wti \tau}_1} > 45\,$GeV, such a signal can be
observed only at future $e^+e^-$ colliders. The only signal LEP~II
is likely to detect is two $\tau$'s +missing energy due to a 
pair production of ${\wti {\tau}}_1$. The two final $\tau$'s are in 
this case more energetic than in the
similar production mechanism in the MSSM, where ${\wti {\tau}}_1$
decays into ${\wti \chi_1}^0 +\tau$ and the neutralino has 
non--negligible mass.  

%
As already observed, an increase in $x$ towards the value one, has 
the effect of decreasing scalar masses, while increasing gaugino 
masses. The parameter $\mu$, now much closer to $M_2$, gives chargino 
and neutralino states much more mixed: the curve relative to the 
heaviest neutralino in Fig.~\ref{spectrum} is lowered to the values 
1.3--1.1 in the shown range of $\Lm$. The ratios of up-- and 
down--squark over gluino masses are now between 0.9 and 1.0. The 
lightest particle of the spectrum is still ${\wti \tau}_1$, 
but also ${\wti e}_R$, ${\wti \mu}_R$ become lighter than 
the lightest neutralino.  

A severe constraint on this model may come from the \bsg 
test. (This was considered in~\cite{DIM,GMSBbsg} in the context 
of more general GMSB models than the MMM.)
The estimate for the branching ratio \br is 
obtained here using the one--loop supersymmetric boundary 
conditions at the electroweak scale given in~\cite{BBMR} and 
including the leading QCD corrections as in~\cite{BBMR,BMMP}. 
%
Experimental errors of some relevant low--energy variables 
as well as theoretical uncertainties in the QCD corrections
are kept into account as in~\cite{BMMP} and~\cite{BDN}. 
The main source of uncertainty comes from the ambiguity in the 
scale at which this process has to be evaluated: we let this 
scale vary between $m_b/2$ to $2 m_b$. 

Among the supersymmetric amplitudes, the collection of those 
relative to chargino exchange deserves a little attention in this 
model. We rewrite them in the form:
\beq
A_{\wti{ \chi}^-} \  = \ C^{\gamma} \sum_{j,k}^{2,6} 
           (- 2) \, x_{wk} \ 
\left \{ 
 C_{1\,jk}^{bs} \Bigl( F_1 +\frac{2}{3} \, F_2\Bigr)(x_{jk}) +
 {m_{\ti \chi_j} \over m_b} \ C_{2\,jk}^{bs} 
     \Bigl( F_3 +\frac{2}{3} \, F_4\Bigr)(x_{jk})
    \right \}
\label{phch}                      
\eeq
where the function $F_i$ are given in~\cite{BBMR}, the symbols
$x_{ij}$ denote ratios of masses: $x_{wk} \equiv M_W^2/m_{\wti u_k}^2$,
$x_{jk} \equiv m_{\wti \chi^+}^2/m_{\wti u_k}^2$, and the constant 
$C^\gamma$ is $C^\gamma \equiv G_F e /(\sqrt{32} \pi^2)$. The 
coefficient $C_{1\,jk}^{bs} $ collects the couplings: 
\beq
C_{1\,jk}^{bs} \ = \ 
         G_{UL}^{jkb}  G_{UL}^{jks*}  
 -\bigl( G_{UL}^{jkb} H_{UL}^{jks*} +H_{UR}^{jkb} G_{UR}^{jks*}\bigr)
 + H_{UR}^{jkb}  H_{UL}^{jks*}  
\eeq
of the ``pure gaugino'', ``mixed gaugino--higgsino'' and 
``pure higgsino'' contributions. By using the definitions 
in~\cite{BBMR} and the fact that $\sin \beta \sim 1$, 
$\cos \beta \sim 1/\tanb$ in this model, $C_{1\,jk}^{bs}$ can be 
re--expressed as:
\beq
C_{1\,jk}^{bs} \ \simeq \
   \vert V_{j1}\vert^2 \Gm_{UL}^{k3} \Gm_{UL}^{*k2} 
 - \sqrt{\frac{x_{tw}}{2}} \,V_{j1} V_{j2}^* 
 \Bigl( \Gm_{UL}^{k3} \Gm_{UR}^{k3} K_{ts} +
        \Gm_{UR}^{k3} \Gm_{UL}^{k2} K_{tb} \Bigr)
 + \frac{x_{tw}}{2} 
 \vert V_{j2}\vert^2 \vert \,\Gm_{UR}^{k3}\vert^2 K_{tb} K_{ts}^* \,,
\label{coeffcuno}
\eeq
where $K$ is the Kobayashi--Maskawa matrix, $U$ and $V$ the matrices
needed to diagonalize the chargino mass matrix. The $3\times 6$ 
matrices $\Gamma_{UL}$ and $\Gamma_{UR}$, juxtaposed, give the 
diagonalization matrix $D_U$ of the up--squark mass matrix. Their 
numerical components as well as those of $\Gm_{DL}$, $\Gm_{DR}$, as 
obtained for $\Lm = 100\,$TeV, are given in Appendix~C. 
Similarly, the coefficient $C_{2\,jk}^{bs}$ multiplied by 
$ {m_{\ti \chi_j} / m_b}$ is:
\beq
 \frac{m_{\ti \chi_j}}{m_b}\, C_{2\,jk}^{bs}
\ \simeq \
 \sqrt{\frac{x_{jw}}{2}} \, \tanb  \,
 \Bigl( 
 - U_{j2} V_{j1}^* \, \Gm_{UL}^{k3} \Gm_{UL}^{*k2} 
 + \sqrt{\frac{x_{tw}}{2}} U_{j2} V_{j2}^* \,
   \Gm_{UL}^{k3} \Gm_{UR}^{*k3} K_{ts}^* \Bigr) \,.
\label{coeffcdue}
\eeq

The amplitude in (\ref{phch}) with the pure gaugino coupling (first 
term in (\ref{coeffcuno})) is the supersymmetric counterpart of the 
SM amplitude:
\beq
A_{SM} \ =  \  C^{\gamma} \  K_{tb} K_{ts}^* \ 
   3 \,x_{tw} \ \Bigl( \frac{2}{3} \, F_1 + F_2 \Bigr)(x_{tw})\,;
\label{phsm}                     
\eeq
those with pure higgsino couplings (the last terms in 
(\ref{coeffcuno}) and (\ref{coeffcdue})), the counterparts 
of the Higgs--mediated amplitude:
\beq
A_{H^-}  \ = \ C^{\gamma} \  K_{ts}^* K_{tb} \
   x_{th} \ \left \{ \cot ^2 \beta \
   \Bigl( \frac{2}{3} \, F_1 + F_2\Bigr)(x_{th})
 + \Bigl( \frac{2}{3} \, F_3 + F_4\Bigr)(x_{th})
   \right \}\,,
\label{phh}   
\eeq
with obvious meaning of all ratios $x_{ij}$. In particular, the 
amplitude with the last coupling in (\ref{coeffcuno}) corresponds to 
the Higgs contribution proportional to $\cot^2\beta$, that with the 
last coupling in (\ref{coeffcdue}) to the $\tanb$--independent 
one. Both couplings have column--index equal to 2 in the 
diagonalization matrices $U$ and $V$, which, indeed, selects the 
higgsino component of the chargino exchanged in the 
loop (see~\cite{HAKA}, whose notation we follow here). We remark that 
in this model the chargino mass matrix,
$X\sim ((M_2, \sqrt{2} M_W ),(0,\mu))$, has 
all positive entries. It is, therefore, $\det X > 0$. The two 
diagonalization matrices $U$, $V$ have then both the structure 
$((\cos\phi_{U,V}, \sin \phi_{U,V}), 
(-\sin\phi_{U,V}, \cos\phi_{U,V}))$ with angles 
$\phi_U$, $\phi_V$ determined by 
$\tan \phi_U \sim  (m_{\wti \chi_1}^2- (M_2^2 + 2 M_W^2))/ 
                     (\sqrt{2} M_W \mu)$ 
and 
$\tan \phi_V \sim (m_{\wti \chi_1}^2 - M_2^2 )/ 
                     (\sqrt{2} M_W M_2)$, both $>0$. 
(We remind that in our convention,  $m_{\wti \chi_1}$ is the 
heavier chargino.)

\begin{figure}[t]
\begin{center} 
\epsfxsize=10.5cm
\leavevmode
\epsfbox[45 390 510 720]{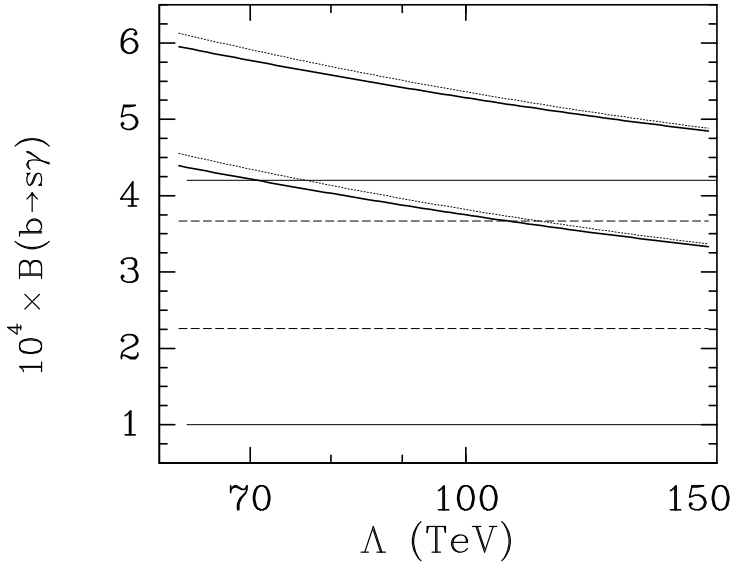}
\end{center}
\caption[f5]{\tenrm{Branching ratio for the decay $b \to s \gamma$ 
obtained in the MMM (solid lines); the dashed lines indicate the
SM's band and the dotted lines the band obtained in a Two-Higgs
doublets model with the same mass for the charged higgs as obtained
in the MMM. The horizontal solid lines delimit the experimentally
allowed values. The results shown correspond to the solutions
$\tanb$--$\mu$ indicated in Fig.~\ref{mmmfigloop} by solid lines.}}
\label{bsg}
\end{figure}

The two amplitudes with couplings (\ref{coeffcdue}) are, in general, 
those responsible for the growth of the chargino contribution for 
increasing $\tanb$ and should, in principle, be the most relevant 
ones when $\tanb$ is large. The situation is, however, slightly 
different in this model. As it can be seen from Appendix~C, the 
$\tanb$--dependent gaugino--higgsino amplitude,
\beq
A_{{\wti W}{\wti H}} \  = \ C^{\gamma} \sum_{j,k}^{2,6} 
            2 \, x_{wk} \,
 \sqrt{\frac{x_{jw}}{2}} \, \tanb  \,
  U_{j2} V_{j1}^* \ \Gm_{UL}^{k3} \Gm_{UL}^{*k2} 
     \Bigl( F_3 +\frac{2}{3} \, F_4\Bigr)(x_{jk}) \,,
\label{mixwh}                      
\eeq
gets contributions 
from the mainly left--handed stop ($k=2$) and the 
mainly left--handed scharm ($k=6$), which have non--negligible 
mixings with ${\wti c}_L$ and ${\wti t}_L$, respectively. 
(The index $k$ is ordered according to increasing values of
the squark masses.) Sizes and signs of these mixings 
are determined by the RGE which generate 
off-diagonal terms in the left--left sector of the up--squark 
mass matrix, and the requirement of orthogonality for $D$.
For our representative value $\Lm = 100\,$TeV, 
$\vert \Gm_{UL}^{23} \Gm_{UL}^{*22}\vert \sim
\vert \Gm_{UL}^{63} \Gm_{UL}^{*62}\vert \sim 0.02$.
The two products have opposite sign and differ only at the
$10^{-6}$ level; the presence of the functions
$(F_3 + 2/3 F_4)(x_{j,k})\sim {\cal O}(1)$, however, 
gives rise to a flavour violation still at the $10^{-3}$ level.
An additional cancellation is due to the gaugino mixing. After
summation in $k$, the sign of the two terms corresponding to the 
two different $j$'s is different, as it appears evident from the 
previous discussion on the matrices $U$ and $V$. (This result, 
obviously does not depend on the freedom used in fixing relative
phases in the elements of $U$ and $V$ (see~\cite{HAKA}).) 
Again, for $\Lm = 100\,$TeV, it is: 
$  U_{12} V_{11}^* \simeq -0.36$, $  U_{22} V_{21}^* \simeq +0.20$, 
($ m_{\wti \chi_2} = 456\,$GeV, $ m_{\wti \chi_1} = 254\,$GeV). 
Furthermore, the ratio $x_{wk}$ is a strong factor of
suppression, $ x_{wk} \sim 10^{-2}$, reducing the 
``gaugino--higgsino'' mixing amplitude (\ref{mixwh})
to be more than two orders of magnitude smaller than
$A_{SM}$ and $A_{H^-}$ and 
roughly one order of magnitude larger than the 
pure ``gaugino'' contribution. 
Similar cancellations appear also in the other gaugino--higgsino 
mixing amplitude (with coupling in the second term of 
(\ref{coeffcuno})), which gives rise to the 
smallest amplitude in the collection (\ref{phch}). 

The $\tanb$--dependent pure ``higgsino'' amplitude gets 
non--negligible contributions from the two stop states ($k=1$ and 
$k=2$). The product of the two elements of 
$\Gamma_{UL}$, $\Gamma_{UR}$, with their opposite sign for the 
two $k$'s, leads to a left--right mixing 
$< 10^{-4}$,
overkilling the enhancement due to $\tanb$.
Numerically, once all summations on $k$ and $j$ are 
performed, this amplitude turns out to be smaller than the 
$\tanb$--independent pure ``higgsino'' amplitude:
\beq
A_{{\wti H}{\wti H}} \  = \ C^{\gamma} \sum_{j,k}^{2,6} 
         (-   2) \, x_{wk} \,
 \frac{x_{tw}}{2} \,
  \vert V_{j2} \vert^2  \vert \Gm_{UR}^{k3}\vert^2 K_{tb} K_{ts}^*
     \Bigl( F_1 +\frac{2}{3} \, F_2\Bigr)(x_{jk}) \,,
\label{pureh}                      
\eeq
where it is $(\Gm_{UR}^{13})^2 \simeq 1$, $V_{22}^2 \simeq 1 $ and 
$x_{tw}/2 \simeq 1 $ and where the flavour violation has the same 
weight as in the SM. The amplitude (\ref{pureh}) is
the largest one in the collection (\ref{phch}).
Its sign is opposite to that of $A_{SM}$ and $A_{H^-}$,
but due to the suppression factor $x_{wk}$, it 
produces only a tiny cancellation of $A_{H^-}$. 

Also the gluino amplitude:
\beq
A_{\wti{g}}  \ =  \  C^{\gamma}  \sum_{k} ^6 
                  \frac{16}{9} \,x_{wk} \
    \Bigl({\al_s \over \al_w} \Bigr)
    \left \{ \Gm_{DL}^{kb} \ \Gm_{DL}x^{*ks} \ F_2(x_{gk}) 
             \ - {m_{\ti g} \over m_b} \
             \Gm_{DR}^{kb} \ \Gm_{DL}^{*ks} \ F_4(x_{gk}) 
    \right \}
\label{phgl}                       \\
\eeq
is known to increase for increasing values of $\tanb$~\cite{IO}. Of 
the two terms in (\ref{phgl}), the larger contribution comes from 
the second one, with exchange of the mainly right--handed sbottom 
($k= 1$), the mainly left-handed sbottom ($k=4$) and the mainly 
left--handed s-strange ($k=6$). Left--right terms in the down--squark 
mass matrix play a larger role than in the up--squark mass matrix and 
the corresponding mass eigenstates are more mixed states. The 
simultaneous left--right and flavour transitions are, however, still 
$\ltap 10^{-4}$. The enhancement factors ${m_{\ti g} / m_b(m_b)}$ and 
${\al_s / \al_w}$, together with the fact that down-- and up--squarks 
have very similar masses, makes this amplitude 
comparable to the $\tanb$--dependent pure ``higgsino'' one. In our 
representative case $\Lm = 100\,$TeV, the SM, Higgs, chargino and 
gluino amplitudes are respectively: $-1.1\times 10^{-9}$, 
$-5.7 \times 10^{-10}$, $ 7.9 \times 10^{-12}$, and  
$1.3 \times 10^{-11}$. The charged Higgs mass, for this value of 
$\Lambda$, is $360\,$GeV. The neutralino amplitude is completely
negligible. 

The values of \br predicted by this model are in the region delimited 
by solid lines in Fig.~\ref{bsg}. The MMM's branching ratio deviates 
little from that obtained in a Two--Higgs--Doublet Model (band 
delimited by dotted lines) with $m_{H^\pm}$ as predicted by the 
MMM. For reference we report also the SM prediction (region within the 
dashed lines) and the experimentally allowed value (region within the 
two horizontal solid lines)~\cite{CLEO}. 

Since the requirement of positivity for the the ${\wti \tau_1}$-mass
has already selected relatively large values of $\Lm$, 
$\Lambda \gtap 62\,$TeV, and the LEP~I constraint 
$m_{\wti \tau_1} \gtap 45\,$GeV imposes $\Lm \gtap 72\,$TeV,
no further exclusion comes, at the moment, from the measurement 
of this decay. Improved experimental results will require 
improvements in the calculation of the 
supersymmetric branching ratio before any significant 
conclusion can be drawn. 
A significant shift of $x$ towards one, however, would push 
already now the MMM's band outside the experimentally
allowed range up to $\Lm \sim 100\,$TeV. For this value of $\Lm$, 
it is: $m_{H^\pm} = 300\,$GeV and the SM, Higgs, chargino and 
gluino amplitudes are: $-1.1\times 10^{-9}$, $-6.7 \times 10^{-10}$, 
$ 4.9 \times 10^{-12}$, and $1.5 \times 10^{-11}$. 

\section{Conclusions}

In this paper, we studied the MMM in the approximation of a 
messenger scale of the same order than 
the supersymmetry--violating messenger scalar mass, and 
gave our estimate for the 
solutions $\tanb$--$\mu$ enforced by the mechanism of radiative 
breaking of $SU(2)_L\times U(1)_Y$. 

Differently than generic GMSB models, which benefit from the 
presence of more free parameters at the messenger scale, 
the MMM requires a careful handling of 
corrections to $B \mu$, $m_{H_u}^2$ and $m_{H_d}^2$, 
which determine the electroweak breaking.
To this aim, we considered the one--loop corrected 
effective potential with contributions coming from all 
massive modes in the model: quark, squark, lepton, slepton, chargino, 
neutralino, gauge and higgs modes. 
Finally we included additional logarithmic two--loop corrections
to the parameter $B$ of same type and size than those induced 
by the one--loop effective potential.
All of these corrections turn out to be 
important to obtain solutions $\tanb$--$\mu$ 
stable under variation of the decoupling scale $Q_0$ around a typical
squark mass. 

Among the predictions obtained by studying the MMM's mass spectrum, 
the most interesting is that ${\wti \tau_1}$ is the lightest 
sparticle. Indeed, the requirement of positivity of 
$m_{{\wti \tau}_1}^2$ as well as that coming from the LEPI lower
bounds on supersymmetric masses, exclude already values of 
$\Lm$ up to $\sim 70\,$TeV. This prediction has other 
important consequences. The MMM, in fact, a) cannot accommodate the 
$ e^+ e^- \gamma \gamma$ CDF event~\cite{CDFPHPH}; b) may be 
detected in $e^+ e^-$ collisions through 
two $\tau$'s +missing energy or four $\tau$'s +missing energy
signals. 
 
As expected, flavour violation effects as well as effects due to 
left--right mixing in the squark sector are small in this model. The 
sparticle contributions to \bsg, opposite to the $W^\pm$ and $H^\pm$ 
contributions are, therefore, not very significant. The overall 
MMM's prediction closely resembles that of a Two Higgs Doublet Model 
with same $H^\pm$ mass. When $m_{\wti \tau_1}$ is larger than 
$45\,$GeV,  we obtain rates consistent with the present experimental 
measurement.

\newpage
{\bf \noindent Acknowledgements \hfill} \\[1.01ex]
I thank many of my colleagues for discussions: M.~Bastero-Gil, 
A.~Brignole, D.~Comelli, M.~Drees, E.~Duchovni, M.~Kugler, 
D.~Lellouch, Y.~Nir, R.~Plesser, N.~Polonsky, R.~Rattazzi, M.~Reuter, 
F.~Vissani, G.~Wolf, and D.~Wyler. I also acknowledge the hospitality 
of the CERN theory group, where part of this work was carried out.

\vskip 0.5truecm
{\bf \noindent Note added \hfill} \\[1.01ex]
After completion of this work we became aware of the existence of 
the paper \hph{97}{01}{341} by D.~Dicus, B.~Dutta, and S.~Nandi which 
studies one of the experimental signals listed in this paper, and of 
the content of \hph{96}{12}{464} by R.~Rattazzi and U.~Sarid with the
same subject of the present paper. The solutions $\tanb$--$\mu$ 
seem in qualitative agreement with ours. Differences appear, however, 
in the conclusions reached for the implication of the measurement of
$\bsg$.

\newpage
\appendix 
\section{}
\noindent
We list the {\bf one--loop RGE} to which we refer in the text. For 
simplicity we give the approximate form valid in the limit of 
Kobayashi--Maskawa matrix $K \sim \bbbone$. The one--loop 
equations actually used in our analysis, with a realistic $K$, 
are given in~\cite{BBMR}.

\noindent 
{\bf -- gauge couplings and gaugino masses}: 
\beq
 \dot{\wal_i} \ = \ -b_i\ \wal_i^2\, \quad \quad \quad 
 \dot{M_i} \ = \ -b_i\ \wal_i M_i 
\label{grge}
\eeq
The dot indicates derivative with respect to 
$t_{XQ} \equiv 2 \ln{(X/Q)}$ and the tilde a division over
$4 \pi$. The supersymmetric $\beta$--function coefficients
$b_i$ are $(-3,1,33/5)$. The corresponding SM ones, $b_i^{SM}$, 
are $(-7,-19/6,41/10)$. 

\noindent 
{\bf -- yukawa couplings and soft trilinear couplings}: \hfill \\
\bea
\dot{\wal}_t                    =  & \!\!\!\!\!\!
 \left(GY\right)_t -6 \wal_t -\wal_b 
       \phantom{-\wal_\tau }       & \quad \quad
\dot{A}_t                       =  
 -\left(GA\right)_t - 6 A_t \wal_t - A_b \wal_b   
 \phantom{- A_\tau \wal_\tau}        \nn  \\
\dot{\wal}_b                    =  & \!\!\!
 \left(GY\right)_b -6 \wal_b -\wal_t -\wal_\tau 
                                   & \quad \quad
\dot{A}_b                       = 
 -\left(GA\right)_b - 6 A_b \wal_b - A_t \wal_t 
 - A_\tau \wal_\tau                  \nn \\    
\dot{\wal}_\tau                 =  & \!\!\!
 \left(GY\right)_\tau -4 \wal_\tau -3 \wal_b 
          \phantom{-\wal_\tau }    & \quad \quad
\dot{A}_\tau                    =  
 -\left(GA\right)_\tau - 4 A_\tau \wal_\tau 
 - 3 A_b \wal_b  \phantom{- A_\tau \wal_\tau}
\label{susyuka}
\eea
where $\alpha_t = h_t^2/ 4 \pi $ and after defining 
$C^Y_i = (13/36),(7/36),(3/4)$ for $i={t,b,\tau}$, 
$GY$ and $GA$ are: 
\bea
 (GY)_i & = & 4\{C_3 \wal_3 +C_2 \wal_2 + \frac{3}{5} C^Y_i \wal_1 \}
                   \nn \\       
 (GA)_i & = & 4\{C_3 \wal_3 M_3 +C_2 \wal_2 M_2 
 + \frac{3}{5} C^Y_i \wal_1 M_1 \} 
\eea 

\noindent 
{\bf -- bilinear coupling}:
\beq
\dot {B} = - 4 \Bigl\{  C_2 \wal_2 M_2
      + \frac{3}{5}\Bigl(\frac{1}{2}\Bigr)^2 \wal_1 M_1
               \Bigr\}
      - ( A_\tau \wal_\tau + 3 A_b \wal_b + 3 A_t \wal_t ) 
\label{brge}
\eeq
{\bf -- first two generations squark and slepton masses }
\beq
\dot{(m_i^2)}_{11,22}                                 = 
 4 \left\{ C_3 \wal_3 M_3^2 + C_2 \wal_2 M_2^2 +
           \frac{3}{5} Y^2 \wal_1 M_1^2 \right \}
\label{twogen}
\eeq

\noindent 
{\bf -- third generation sfermion masses}:
\bea
\dot{(m_{\wti Q}^2)}_{33}  &\!\!\!  = &\!\!\! 
\dot{(m_{\wti Q}^2)}_{11,22}
     - \phantom{2}\ \wal_t  \left(SS\right)_t
     -              \wal_b  \left(SS\right)_b   \nn \\  
\dot{(m_{\wti U}^2)}_{33}  &\!\!\!  = &\!\!\! 
\dot{(m_{\wti U}^2)}_{11,22}
     -2\ \wal_t \left(SS\right)_t \nn 
\quad \quad \quad  \quad \quad \quad \ 
\dot{(m_{\wti D}^2)}_{33}    = \dot{(m_{\wti D}^2)}_{11,22}
     -2\ \wal_b \left(SS\right)_b              \nn \\             
\dot{(m_{\wti L}^2)}_{33}  &\!\!\!  = &\!\!\! 
\dot{(m_{\wti L}^2)}_{11,22}
  - \phantom{2}\ \wal_\tau  \left(SS\right)_\tau \,           
\quad \quad \quad  \quad \quad \quad  
\dot{(m_{\wti E}^2)}_{33}    =  \dot{(m_{\wti E}^2)}_{11,22}
     -2\ \wal_\tau  \left(SS\right)_\tau\ .
\label{thirdgen}
\eea
\noindent 
{\bf -- soft Higgs-potential parameters}:
\bea
\dot{(m_{H_d}^2)} &\!\!\! = &\!\!\! \dot{(m_{\wti L}^2)}_{11,22}
 - 3 \wal_b \left(SS\right)_b  - \wal_\tau \left(SS\right)_\tau\,; 
\quad \quad \quad 
\dot{(m_{H_u}^2)}         =         \dot{(m_{\wti L}^2)}_{11,22}
  - 3 \wal_t  \left(SS\right)_t
\label{m12}
\label{higgsrge}
\eea
where 
\beq
\begin{array}{lcl}
\left(SS\right)_t     & = & 
 (m_{\wti Q}^2 +m_{\wti U}^2)_{33} + m_{H_u}^2 + A_t^2  
                                                \nn  \\[1.4ex]
\left(SS\right)_b     & = & 
 (m_{\wti Q}^2 +m_{\wti D}^2)_{33} + m_{H_d}^2 + A_b^2         
                                                \nn  \\[1.4ex]
\left(SS\right)_\tau  & = & 
 (m_{\wti L}^2 +m_{\wti E}^2)_{33} + m_{H_d}^2 + A_\tau^2 \,.
\end{array}
\label{sspar}
\eeq
\noindent 
{\bf -- $\mu$ parameter}:
\beq 
\dot{(\mu^2)} \ = \ 
 \left[4\Bigl( C_2 \wal_2
 + \frac{3}{5} \Bigl(\frac{1}{2}\Bigr)^2 \wal_1\Bigr) 
 - \left( 3 \wal_t + 3 \wal_b + \wal_\tau \right)
 \right] \mu^2 
\eeq
The low--energy parameter $\mu$ is obtained from the 
minimization condition of the scalar Higgs potential. This 
equation may help to trace back the high--energy value of 
this parameter.

We report also the additional {\bf two--loop} terms in the RGE for 
$B$ to which we refer in the text (see~\cite{MAVAU}):
\bea 
\dot {B}
 \!\! &\!\! =\!\!&  \!\! \phantom{+12}
   \left\{ \frac{80}{3} \left( C_2\, \wal_2\right)^2 M_2
          + \frac{12}{5} \left(C_2\,\wal_2\right)
              \left(\wal_1\right) \left(M_1+M_2\right)  
          + \frac{207}{25} \left(\wal_1\right)^2  M_1)
    \right\}                                     \nn\\[1.01ex]
 \!\! &\!\! \!\!&  \!\!               
 + 12 \left\{
 \left[C_3 \,\wal_3 M_3 + \frac{1}{15}\wal_1 M_1 \!
 \right] \wal_t +    
 \left[C_3 \,\wal_3 M_3 - \frac{1}{30}\wal_1 M_1 \!
 \right] \wal_b 
 +   \frac{1}{10}\wal_1 M_1  \wal_\tau   
      \right\}                                   \nn\\[1.01ex]
 \!\! &\!\! \!\!& \!\!               
 - 12 \left\{
 \left[C_3 \,\wal_3  +\frac{1}{15}\wal_1 \! \right] A_t \wal_t
  \ +  \     
 \left[C_3 \,\wal_3  -\frac{1}{30}\wal_1 \! \right] A_b \wal_b
  \ +  \    
 \frac{1}{10}\wal_1   A_\tau  \wal_\tau 
      \right\}                                   \nn\\[1.02ex]
 \! &\!\! \!\!& \!
 + \phantom{1} 6 \
  \left[ A_\tau \wal_\tau^2  + 
          \left( A_t + A_b\right) \wal_t \wal_b 
        + 3 A_t \wal_t^2  + 3 A_b \wal_b^2 
  \right] \,.
\label{BRGEII}
\eea

\section{}
It is easy to see that, if we neglect phases other than that in the 
CKM matrix, the scalar potential $V$ (here indifferently $V_0$ or 
the fully one--loop corrected potential $V_1$) 
depend on the real components $\phi_i,\psi_i$ ($i=u,d$)
of the two neutral higgses $H_d^0$ and $H_u^0$
($ H_d^0  \, \equiv\, \phi_d + i \psi_d\,, 
   H_u^0  \, \equiv\, \phi_u + i \psi_u $)~\footnote{The 
 physical neutral Higgs field $H^0$, $h^0$, $A^0$, properly normalized, 
 are expressed in terms of $H_d^0$ and $H_u^0$ as in~\cite{GUHA}}
through the bilinear operators 
\beq
 \epsilon_d  \equiv \vert H_d^0 \vert^2\,;  \quad \quad 
 \epsilon_u  \equiv \vert H_u^0 \vert^2\,;  \quad \quad   
 \epsilon_3  \equiv  H_u^0 H_d^0 + h.c. \,.
\label{combin}
\eeq 
(If extra phases are allowed, one would have to consider slightly 
more general operators. The following results would remain unchanged 
up to complex conjugates of soft parameters and/or $\mu$).

The derivatives of $V$ with respect to $\phi_i$, which determine the 
minimum conditions, can then be expressed in terms of derivatives 
with respect to $\epsilon_d$, $\epsilon_u$ and $\epsilon_3$ as 
follows:
\bea
 (\spd{\phi_d} V )                                       & = & 
 {2 \phi_d} \,(\spd{\epsilon_d} V)                         +
 {2 \phi_u} \,(\spd{\epsilon_3} V)                       \nn \\
 (\spd{\phi_u} V)                                        & = & 
 {2 \phi_u} \,(\spd{\epsilon_u} V)                         + 
 {2 \phi_d} \,(\spd{\epsilon_3} V)\,.        
\label{deriveven}
\eea
The derivatives $\diii V$ are closely related to the entries of 
the pseudoscalar Higgs mass matrix, if one imposes that 
$\langle \psi_d \rangle = \langle \psi_u \rangle= 0$: 
${\partial^2 V}/{\partial \psi_d^2} =  2 \duno V;$     
${\partial^2 V}/{\partial \psi_d \partial \psi_u} = -2 \dtre V;$
${\partial^2 V}/{\partial \psi_u^2} =  2 \ddue V $.
Moreover, at the electroweak minimum, i.e. when 
$ \spd{\phi_d} V =  \spd{\phi_u} V = 0$ for
$\langle \phi_d \rangle = v_d $, $\langle \phi_u \rangle = v_u$ 
($\tanb \equiv {v_{u}}/{v_{d}}$), 
it is:
\beq
 \duno V = - \dtre V \tanb; \quad \quad 
 \ddue V = - \dtre V \cot \beta \,.
\eeq 
In other words, the correct minimization condition guarantees the 
existence of one Goldstone boson mode, corresponding to the 
eigenstate $ (\cos\beta, -\sin \beta)$. The massive mode
$ (\sin \beta, \cos \beta)$ has mass 
$ (2/\sin 2\beta ) (-\dtre V)$. In this sense 
$-\dtre V$ identifies $ (B\mu)$ when $V$ is $V_0$ and what will be 
called ${\widehat {B \mu}}$ in the case of $V_1$.

The derivatives 
$\spd{\epsilon_u}$, $\spd{\epsilon_d}$, and
$\spd{\epsilon_3}$ of the RGE--improved tree--level potential, 
are: 
\bea
\spd{\epsilon_d}{V_0} &=& m^2_{H_d} + \mu^2 + \ggg \epsm /4
\nn\\
\spd{\epsilon_u}{V_0} &=& m^2_{H_u} + \mu^2 - \ggg \epsm /4 
\nn\\
\spd{\epsilon_3}{V_0} &=& -\left(B \mu\right) \,,
\eea 
with $\epsilon_{\pm} \equiv \epsilon_d \pm \epsilon_u$. Those
of the one--loop corrected potential $V_1$ 
maintain the same form when expressed 
in terms of ``hatted'' variables ${\widehat m}^2_{H_d}$, 
${\widehat m}^2_{H_u}$, and ${\widehat{ B \mu}}$, where 
${\widehat m}^2_{H_i} = m^2_{H_i} + \delta m^2_{H_i}$, 
and ${\widehat{ B \mu}} = B \mu + \delta (B\mu)$. The shifts 
are due to the one--loop corrections coming from the following 
sectors:
quark/squark, lepton/slepton, 
chargino/charged gauge and Higgs mode, neutralino/neutral
gauge and Higgs mode. They can be evaluated generalizing 
methods used in~\cite{ZWIRN1,MANUEL}. 
We list them in the following. 

\noindent {\bf top--stop contribution} 

\noindent 
\bea 
 {\delta m}^2_{H_d} \vert_{t,\,{\wti t}}    \!\!\! & = & \!\!\!
  \frac{3}{32\pi^2} \left[
  S(m_{{\wti t_1},{\wti t_2}}) \frac{\ggg}{8}  +
  D(m_{{\wti t_1},{\wti t_2}}) \left( \left(h_t \mu\right)^2 \!+\!   
   \frac{\ggg}{8} \, C_{\wti t} \, \Delta^{2}_{\wti t} \right)
                    \right]               \nn \\
 {\delta m}^2_{H_u} \vert_{t,\,{\wti t}}    \!\!\! & = & \!\!\!
  \frac{3}{32\pi^2} \left[
  S(m_{{\wti t_1},{\wti t_2}})
                       \left( \! h_t^2 \!-\!\frac{\ggg}{8}\!\right) +
  D(m_{{\wti t_1},{\wti t_2}}) \left(\!\left(h_t A_t \right)^2  \!-\!   
   \frac{\ggg}{8} \, C_{\wti t} \, \Delta^{2}_{\wti t} \right)
            \! - \! 2 F(m_t,Q_0)                
                    \right]                \nn  \\      
 {\delta (B\mu) } \vert_{t,\,{\wti t}}    \!\!\! & = & \!\!\! 
  \frac{3}{32\pi^2} \left[ 
      D(m_{{\wti t_1},{\wti t_2}}) 
              \left(A_t\, \mu \, h_t^2 \right)\right] \,, 
\label{topstop}
\eea
where $C_{\wti t}$ is 
$ C_{\wti t} \equiv (1\! -\!\frac{8}{3}\sin^2\theta_W ) $ 
and $ \Delta^{2}_{\wti t}$ is defined as 
$ \Delta^{2}_{\wti t} \equiv 
 (m_{\wti Q}^2 -m_{\wti U}^2)_{33} + \epsm\, C_{\wti t}\, \ggg/4 $. 
The functions $S(m_{{\wti t_1},{\wti t_2}})$,
$D(m_{{\wti t_1},{\wti t_2}})$, 
where the dependence on the scale $Q_0$ has been 
suppressed for compactness, can be expressed in terms of 
$ F(m,Q_0) \equiv  2\, m^2 \left[ \log({m^2}/{Q_0^2}) -1 \right]$ 
as:
\bea
 S(m_{{\wti t_1},{\wti t_2}})        & \equiv &  \ \ 
      F(m_{\wti t_1},Q_0) + F(m_{\wti t_2},Q_0)               \nn \\
 D(m_{{\wti t_1},{\wti t_2}})        & \equiv & 
  \left( F(m_{\wti t_1},Q_0) - F(m_{\wti t_2},Q_0)\right) / 
  \left( m^2_{\wti t_1} - m^2_{\wti t_2} \right)\,.
\eea
The approximation of $2 \times 2$ mass matrices for each generation 
of sfermions is used here, in spite of the fact that the building 
blocks for these matrices are determined without neglecting 
intergenerational mixing terms. All the corrections listed in this 
appendix are clearly not sensitive to these tiny effects. 

\noindent {\bf bottom--sbottom contribution} 
\bea 
 {\delta m}^2_{H_d} \vert_{b,\,{\wti b}}            \!\!\! & = & \!\!\!
  \frac{3}{32\pi^2} \left[
  S(m_{{\wti b_1},{\wti b_2}}) \left(h_b^2 \!-\!\frac{\ggg}{8}\!\right) +
  D(m_{{\wti b_1},{\wti b_2}}) \left(\! \left(h_b A_b \right)^2\! -\!   
   \frac{\ggg}{8} \, C_{\wti b} \, \Delta^{2}_{\wti b} \right)
              - 2 F(m_b,Q_0)                
                    \right]                \nn  \\      
 {\delta m}^2_{H_u} \vert_{b,\,{\wti b}}            \!\!\! & = & \!\!\!
  \frac{3}{32\pi^2} \left[
  S(m_{{\wti b_1},{\wti b_2}}) \frac{\ggg}{8} +
  D(m_{{\wti b_1},{\wti b_2}}) \left( \left(h_b \mu\right)^2 +   
   \frac{\ggg}{8} \, C_{\wti b} \, \Delta^{2}_{\wti b} \right)
                    \right]               \nn \\
 {\delta (B\mu) } \vert_{b,\,{\wti b}}            \!\!\! & = & \!\!\! 
  \frac{3}{32\pi^2} \left[ 
      D(m_{{\wti b_1},{\wti b_2}})
            \left(A_b\, \mu \, h_b^2 \right)\right] \,, 
\label{botsbot}
\eea
with 
$ C_{\wti b} \equiv (1\! -\!\frac{4}{3}\sin^2\theta_W ) $
and $ \Delta^{2}_{\wti b} $ now defined as 
$  \Delta^{2}_{\wti b} \equiv 
(m_{\wti Q}^2 -m_{\wti D}^2)_{33} - \epsm\, C_{\wti b}\, \ggg/4$.

\noindent {\bf sneutrino contribution} 

\noindent 
For each sneutrino species, we have:
\bea 
 {\delta m}^2_{H_d} \vert_{{\wti \nu}}    \!\!\! & = & \!\!\!
  \frac{1}{32\pi^2} \left[\
  F(m_{\wti \nu},Q_0) \,\frac{\ggg}{8}  
                    \right]               \nn \\
 {\delta m}^2_{H_u} \vert_{{\wti \nu}}    \!\!\! & = & \!\!\!
  \frac{1}{32\pi^2} \left[ -
  F(m_{\wti \nu},Q_0) \frac{\ggg}{8}  
                    \right]                \nn  \\      
 {\delta (B\mu) } \vert_{{\wti \nu}}    \!\!\! & = & \ 0\,.
\label{sneutr}
\eea

\noindent {\bf tau--stau contribution} 

\noindent 
It can be obtained from the bottom-sbottom contribution after 
suppression of the colour factor 3, with obvious
replacements, i.e. $h_b\to h_\tau$, $m_b\to m_\tau$, 
$m_{\wti b_i} \to m_{\wti \tau_i}$, 
$C_{\wti b}\to C_{\wti \tau} \equiv (1-4 \sin^2\theta_W )$, and 
$  \Delta^{2}_{\wti b} \to \Delta^{2}_{\wti \tau} \equiv 
(m^2_L -m^2_E)_{33} - \epsm\, C_{\wti \tau}\, \ggg/4$.

\noindent {\bf first--two--generations quark/squark, lepton/slepton
 contribution} 

\noindent 
Since the approximation of massless first--two--generations quarks
and leptons is used in this analysis, this contribution can 
be obtained from (\ref{topstop},\ref{botsbot}) and the 
tau-stau contribution putting to zero yukawa couplings,
quark and lepton masses, and substituting third generation sfermion 
masses with the corresponding first and second generation ones.
This contribution is rather small:
there are no corrections coming from this sector to $B\mu$, whereas the 
corrections to $ m^2_{H_u} $ and $ m^2_{H_d} $ from the two 
component of a $SU(2)_L$ doublet
supermultiplet, although singularly not small, 
tend to cancel each other without the diversifying effect of the  
yukawa couplings. 

\noindent {\bf chargino/charged boson contribution} 

\noindent 
The spin $1/2$ modes, charginos, contribute as: 
\bea 
 {\delta m}^2_{H_d} \vert_{{\wti \chi}^\pm}      \!\!\! & = & \!\!\!
  \frac{1}{32\pi^2} \,g^2\left[
 - S(m_{{\wti \chi_1},{\wti \chi_2}}) 
 - D(m_{{\wti \chi_1},{\wti \chi_2}})
            \left( M_2^2+\mu^2 +g^2 \epsm \right) 
                    \right]               \nn \\
 {\delta m}^2_{H_u} \vert_{{\wti \chi}^\pm}       \!\!\! & = & \!\!\!
  \frac{1}{32\pi^2} \,g^2\left[
 - S(m_{{\wti \chi_1},{\wti \chi_2}}) 
 - D(m_{{\wti \chi_1},{\wti \chi_2}})
      \left( M_2^2+\mu^2 -g^2 \epsm \right) 
                    \right]                \nn  \\      
 {\delta (B\mu) } \vert_{{\wti \chi}^\pm}    \!\!\! & = & \!\!\! 
  \frac{1}{32\pi^2} \,g^2\left[
 \ D(m_{{\wti \chi_1},{\wti \chi_2}}) (2\,M_2\mu) \right] \,, 
\label{charg}
\eea
whereas the charged gauge boson mode gives:
\bea 
 {\delta m}^2_{H_d} \vert_{{W}^\pm}             \!\!\! & = & \!\!\!
  \frac{3}{32\pi^2} \left[
   F(M_W,Q_0) \,\frac{\,g^2}{2} 
                    \right]        \nn \\[1.01ex]
 {\delta m}^2_{H_u} \vert_{{W}^\pm}             \!\!\! & = & \!\!\!
  \frac{3}{32\pi^2} \left[
   F(M_W,Q_0) \,\frac{\,g^2}{2}
                    \right]          \nn  \\      
 {\delta (B\mu) } \vert_{{W}^\pm} \!\!\! & = & \ 0\,,
\label{wmode}
\eea
with $M_W$ defined as $M_W = \epsp g^2/2  $.
There are also two charged Higgs modes $ H^\pm,G^\pm $
with ``masses'' given by  
$$
 m^2_{H^\pm,G^\pm} = 
 \frac{1}{2}\left\{\!
 \mu^2_{H_d}\! +\!\mu^2_{H_u} \! +\! \frac{g^2}{2} \epsp
 \pm \left[\left( \mu^2_{H_d}\! -\!\mu^2_{H_u} \!+ 
    \!\frac{g'^2}{2}  \epsm   \right)^2 + 
  \left(2 (B\mu) \!+\! \frac{g^2}{2} \epsilon_3 \right)^2 
     \right]^{\frac{1}{2}} \right\} \,,
$$
with $ \mu^2_{H_i} \equiv m^2_{H_i}+\mu^2$. 
At the $V_0$-- minimum, where the relations
$ \mu^2_{H_d} + \ggg (v_d^2 \!-\! v_u^2) / 4 = (B\mu) \tanb$ and 
$ \mu^2_{H_u} - \ggg (v_d^2 \!-\! v_u^2) / 4 = (B\mu) \cot \beta$ 
hold, $G^\pm$ is the charged Goldstone boson, $H^\pm$ the usual 
charged Higgs, with mass 
$ m^2_{H^\pm} =  \mu^2_{H_u}+ \mu^2_{H_d} + M^2_W$. Both 
modes, however, contribute to the determination of the 
$V_1$--minimum (where the charged Goldstone boson, as well as the
neutral one mentioned below, are not massless anymore). They shift 
the variables $ m^2_{H_u}$, $ m^2_{H_d}$ and $ B\mu$ by the 
quantities: 
\bea 
 {\delta m}^2_{H_d} \vert_{H^\pm,G^\pm}   \!\!\! & = & \!\!\!
  \frac{1}{32\pi^2}  \left[
 \ S({ m_{H^\pm,G^\pm}}) \,\frac{g^2}{4}  
 + D({ m_{H^\pm,G^\pm}}) \,\frac{g'^2}{4}
  \left( m^2_{H_d} -m^2_{H_u} + \frac{g'^2}{2} \epsm \right) 
                    \right]               \nn \\
 {\delta m}^2_{H_u} \vert_{H^\pm,G^\pm}   \!\!\! & = & \!\!\!
  \frac{1}{32\pi^2}  \left[
 \  S({ m_{H^\pm,G^\pm}}) \, \frac{g^2}{4} 
 -  D({ m_{H^\pm,G^\pm}}) \, \frac{g'^2}{4}
  \left( m^2_{H_d} -m^2_{H_u} + \frac{g'^2}{2} \epsm \right) 
                    \right]                \nn  \\      
 {\delta (B\mu) } \vert_{H^\pm,G^\pm} \!\!\! & = & \!\!\! 
  \frac{1}{32\pi^2}  \left[
  \,  - D({ m_{H^\pm,G^\pm}}) \, \frac{g^2}{4} 
  \left(2 (B\mu) + \frac{g^2}{2} \epsilon_3  \right)  \right] \,. 
\label{hmode}
\eea

\noindent {\bf neutralino/neutral boson contribution} 

\noindent 
In this sector, there are: i) one
gauge boson mode, with ``mass'' $M_Z\equiv \epsp \ggg/2 $, which  
contribute as:
\bea 
 {\delta m}^2_{H_d} \vert_{{Z}}      \!\!\! & = & \!\!\!
  \frac{3}{32\pi^2} \left[
  F(M_Z,Q_0)\,\frac{\ggg}{4} 
                    \right]       \nn \\[1.01ex]
 {\delta m}^2_{H_u} \vert_{{Z}}       \!\!\! & = & \!\!\!
  \frac{3}{32\pi^2} \left[
  F(M_Z,Q_0)\,\frac{\ggg}{4}
                    \right]       \nn  \\      
 {\delta (B\mu) } \vert_{{Z}}   \!\!\! & = &  \ 0 \,;
\label{zmode}
\eea
ii) 
two neutral pseudoscalar Higgs modes $A^0,G^0$ 
 with ``masses'' 
$$
m^2_{A^0,{G^0}} = 
 \frac{1}{2}\left\{\!
  \mu^2_{H_d}\! +\!\mu^2_{H_u}   \pm 
 \left[\left( \mu^2_{H_d}\! -\!\mu^2_{H_u} \!+\! \frac{\ggg}{2} \epsm
       \right)^2 + 
  \left(2 (B\mu) \right)^2 
     \right]^{\frac{1}{2}} \right\} 
$$
At the $V_0$--minimum $G^0$ is massless, $A^0$ is the usual 
pseudoscalar Higgs with mass 
$m^2_{A^0} = m^2_{H_d} + m^2_{H_u} + 2 \mu^2 $. They 
contribute as:
\bea
 {\delta m}^2_{H_d} \vert_{{A^0,G^0}}      \!\!\! & = & \!\!\!
  \frac{1}{32\pi^2} \frac{\ggg}{8} \left[
 \ D({ m_{A^0,{G^0}}}) \left( m^2_{H_d} -m^2_{H_u}
           + \frac{\ggg}{2} \epsm \right) 
                    \right]               \nn \\
 {\delta m}^2_{H_u} \vert_{{A^0,G^0}}      \!\!\! & = & \!\!\!
  \frac{1}{32\pi^2} \frac{\ggg}{8} \left[
 - D({ m_{A^0,{G^0}}}) \left( m^2_{H_d} -m^2_{H_u}
           + \frac{\ggg}{2} \epsm \right) 
                    \right]                \nn  \\      
 {\delta (B\mu) } \vert_{{A^0,G^0}} \!\!\! & = & \ 0\,;
\label{namode}
\eea
iii) two neutral scalar Higgs modes with ``masses'' 
$$
m^2_{H^0,h^0} = 
 \frac{1}{2}\left\{\!
  \mu^2_{H_d}\!\! +\!\mu^2_{H_u} \! +\! \frac{\ggg}{2} \epsp
 \pm \left[  \left( \mu^2_{H_d}\!\! -\!\mu^2_{H_u} \!+ \!\ggg \epsm
           \right)^{\!2} \!\!+ \!\!
  \left(\!2 (B\mu) \!+\! \frac{\ggg}{2} \epsilon_3\! \right)^{\!\!2} 
    \! \right]^{\!\frac{1}{2}} \!\right\} 
$$
which, at the $V_0$--minimum, reduce to the conventional 
neutral Higgs masses
$m^2_{H^0,h^0} = 
\{  \mu^2_{H_d}\!\! +\!\mu^2_{H_u} \! +\! M^2_Z    \pm  
 [
 ( \mu^2_{H_d}\!\! +\!\mu^2_{H_u} \! -\! M^2_Z)^2
 \cos^2 2 \beta + 
 ( \mu^2_{H_d}\!\! +\!\mu^2_{H_u} \! +\! M^2_Z)^2
 \sin^2 2 \beta ]^{1/2}\} / 2
 $. 
They produce the shifts:
\bea 
 {\delta m}^2_{H_d} \vert_{{H^0,h^0}}            \!\!\! & = & \!\!\!
  \frac{1}{32\pi^2} \frac{\ggg}{4} \left[
   \frac{1}{2} S({ m_{H^0,h^0}}) 
 + D({ m_{H^0,h^0}}) \left( m^2_{H_d} -m^2_{H_u}
           + \ggg\epsm \right) 
                    \right]               \nn \\
 {\delta m}^2_{H_u} \vert_{{H^0,h^0}}           \!\!\! & = & \!\!\!
  \frac{1}{32\pi^2} \frac{\ggg}{4} \left[
   \frac{1}{2} S({ m_{H^0,h^0}}) 
 - D({ m_{H^0,h^0}}) \left( m^2_{H_d} -m^2_{H_u}
           + \ggg\epsm \right) 
                    \right]                \nn  \\      
 {\delta (B\mu) } \vert_{{H^0,h^0}}  \!\!\! & = & \!\!\! 
  \frac{1}{32\pi^2} \frac{\ggg}{4} \left[
  \,  - D({ m_{H^0,h^0}}) 
      \left( (B\mu) + \frac{\ggg}{4} \epsilon_3
                         \right)  \right] \,; 
\label{nhmode}
\eea
iv) four spin $1/2$ modes, the neutralinos. We find their contribution
following~\cite{ZWIRN2,VIENNA,DENNIS}:
\bea
 {\delta m}^2_{H_d} \vert_{{\wti \chi}^0}        \!\!\! & = & \!\!\!
  \frac{1}{32\pi^2} \left[
  - 2 \sum_{i=1}^4  F(m_{{\wti \chi_i}^0},Q_0) 
        \,  \duno m^2_{{\wti \chi_i}^0}  
                    \right]                \nn  \\      
 {\delta m}^2_{H_u} \vert_{{\wti \chi}^0}        \!\!\! & = & \!\!\!
  \frac{1}{32\pi^2} \left[
  - 2 \sum_{i=1}^4  F(m_{{\wti \chi_i}^0},Q_0) 
        \,  \ddue m^2_{{\wti \chi_i}^0}  
                    \right]                \nn  \\      
 {\delta (B\mu) } \vert_{{\wti \chi}^0}         \!\!\! & = & \!\!\! 
  \frac{1}{32\pi^2} \left[
  + 2 \sum_{i=1}^4  F(m_{{\wti \chi_i}^0},Q_0) 
        \,  \dtre m^2_{{\wti \chi_i}^0}  
                    \right]       
\label{neutralino}
\eea
where in turn, the derivatives of the masses are:
\beq
\diii m^2_{\chi_i^0}   =    - \, 
 \frac{\ m_{{\wti \chi_i}^0}^6\,(\diii c_2)  + 
         m_{{\wti \chi_i}^0}^4\,(\diii c_4)  + 
         m_{{\wti \chi_i}^0}^2\,(\diii c_6)  + (\diii c_8)\ } 
 {\ c_6 + 2 c_4 \,m_{{\wti \chi_i}^0}^2 + 
          3 c_2 \,m_{{\wti \chi_i}^0}^4 
        + 4 c_0 \,m_{{\wti \chi_i}^0}^6\ }  \quad 
 (\epsilon_i= \epsilon_d, \epsilon_u, \epsilon_3) \,.
\eeq
The coefficients $c_i$ are the coefficients of the 
characteristic equation 
$ c_0 \,m_{\chi_i^0}^8 + c_2 \,m_{\chi_i^0}^6 + 
  c_4 \,m_{\chi_i^0}^4 + c_6 \,m_{\chi_i^0}^2 + c_8 = 0 $, 
for the matrix 
${\cal M}_{{\wti \chi}^0} {\cal M}_{{\wti \chi}^0}^\dagger $, 
with all Higgs dependences reinstated back in 
${\cal M}_{{\wti \chi}^0}$. (The fact that at the true minimum 
charged Higgses do not develop vacuum expectation values is 
already imposed, allowing therefore the chargino and neutralino 
mass matrices to decouple.) They are : 
\bea 
      c_0 &=& + 1 
\nn  \\[1.01ex]
      c_2 &=& - 2  \mu^2 - M_1^2 - M_2^2 
              -  \ggg \epsp
\nn  \\[1.01ex]            
      c_4 &=&  + \mu^4 + 2  \mu^2 (M_1^2 +M_2^2) + M_1^2 M_2^2 
               +   [ \mu^2 \ggg +(M_2^2 g'^2+M_1^2 g^2)]
                  \epsp
\nn  \\
          & & 
              -   \mu (M_1 g'^2+M_2 g^2)  \epsilon_3 
              +  \ggg^2  \epsp^2 /4
\nn  \\[1.01ex]       
      c_6 &=& -\mu^4 (M_1^2+M_2^2) - 2  \mu^2 M_1^2 M_2^2 
              +   \mu [\mu^2 (M_1 g'^2+M_2 g^2)
                       +M_1 M_2 (M_2 g'^2+M_1 g^2) ]\epsilon_3
\nn  \\
          & & 
         -   \mu^2 \ggg^2  \epsilon_d \epsilon_u 
         -   \mu^2 (M_2^2 g'^2+M_1^2 g^2)  \epsp
         - (M_2 g'^2+M_1 g^2)^2 \epsp^2  /4
\nn  \\[1.01ex]      
      c_8 &=& + \mu^2 [ \mu^2 M_1^2 M_2^2 
         -   \mu M_1 M_2 (M_2 g'^2+M_1 g^2) \epsilon_3 
         +   (M_2 g'^2+M_1 g^2)^2  \epsilon_d \epsilon_u
                     ] \,.
\eea      

\section{}
We collect here the numerical entries of the up-- and down--squark
mass matrices, their eigenvalues and eigenvectors, as obtained at 
$\Lm = 100\,$TeV. For better readability, we write here ``linear''
mass matrices: their elements are the square root of the entries 
in the real mass matrix; the reported signs 
are those of the corresponding squared elements. The up--squark
mass matrix, split in the four left--left, left--right, right--left
and right--right submatrices is: 
\beq
{\wti M}_U = 
\left(
\begin{tabular}{rrr|rrr}
1099.432 &    0.000 &  -25.320 &    0.000 &    0.000 & -3.879 \\
   0.000 & 1099.432 &   53.712 &    0.000 &    0.000 &  8.228 \\
 -25.320 &   53.712 & 1044.930 &    0.000 &    0.000 & 38.766 \\ \hline
   0.000 &    0.000 &    0.000 & 1054.342 &    0.000 &  0.000 \\
   0.000 &    0.000 &    0.000 &    0.000 & 1054.342 &  0.000 \\
-3.879   & 8.228    &  -38.766 &    0.000 &    0.000 & 976.915
\end{tabular} 
\right) \,.
\label{umm}
\eeq
As in~\cite{BBMR}, we work in a quark basis in which the transition 
from current to mass eigenstates is obtained through the rotation 
of the left--handed up quarks only. The rotation matrix is the 
Kobayashi--Maskawa matrix. The effect of this matrix on both 
up-- and down--squark mass matrices 
is kept up to ${\cal O}(\lambda^3)$, where $\lambda$ is the 
Cabibbo angle. 

The diagonalization matrix for ${\wti M}_u$ is:
\beq
D_U = \left( \begin{tabular}{rrr|rrr}
    .0001 & -.0004 &  .0108 &   .0000 &  .0000 &  .9999 \\
    .0053 & -.0238 &  .9996 &   .0000 &  .0000 & -.0108 \\
    .0000 &  .0000 &  .0000 &   .0369 &  .9993 &  .0000 \\
    .0000 &  .0000 &  .0000 &  -.9993 &  .0369 &  .0000 \\
   -.9762 & -.2169 &  .0000 &   .0000 &  .0000 &  .0000 \\
   -.2169 &  .9759 &  .0244 &   .0000 &  .0000 &  .0001   
\end{tabular} \right) \hspace*{0.2cm}
\begin{array}{r}
 k = 1 \\  2 \\  3 \\  4 \\  5 \\   6 
\end{array} \hspace*{0.3cm}
\left(\begin{array}{r}
   976.907 \\
  1044.902 \\
  1054.342 \\
  1054.342 \\
  1099.432 \\
  1099.466 
\end{array}\right) \,.
\label{diagup}
\eeq  
Each row $k$ represents the decomposition of the $k$--eigenvector 
over the initial basis 
$({\wti u}_L,{\wti c}_L,{\wti t}_L,{\wti u}_R,
{\wti c}_R,{\wti t}_R)$. 
The $6\times 3$ matrices on the left and right of the vertical 
bar are, then, nothing else but the matrices $\Gm_{UL} $ 
and $\Gm_{UR}$ used in the text. The last column of numbers in 
(\ref{diagup}) gives for each $k$, the corresponding eigenvalue
in $GeV$. 

The mass matrix for the down--squark sector looks like:
\beq
{\wti M}_D = 
\left(
\begin{tabular}{rrr|rrr}
1102.331 &   0.000 & -30.548 &   0.000 &   0.000 &   0.000  \\   
   0.000 &1102.331 &  64.803 &   0.000 &   0.000 &   0.000  \\ 
 -30.548 & 64.803  &1033.967 &   0.000 &   0.000 &-256.273  \\ \hline
   0.000 &   0.000 &   0.000 &1050.701 &   0.000 &   0.000  \\ 
   0.000 &   0.000 &   0.000 &   0.000 &1050.701 &   0.000  \\ 
   0.000 &   0.000 &-256.273 &   0.000 &   0.000 & 999.164     
\end{tabular} 
\right) 
\label{dmm}
\eeq
and the corresponding diagonalization matrix is:
 \beq
D_D = \left( \begin{tabular}{rrr|rrr}
    .0018 &  -.0081 &   .5101 &    .0000 &   .0000 &   .8601 \\  
    .0000 &   .0000 &   .0000 &    .9998 &   .0211 &   .0000 \\  
    .0000 &   .0000 &   .0000 &   -.0211 &   .9998 &   .0000 \\  
   -.0073 &   .0329 &  -.8595 &    .0000 &   .0000 &   .5100 \\  
   -.9762 &  -.2169 &   .0000 &    .0000 &   .0000 &   .0000 \\  
    .2168 &  -.9756 &  -.0332 &    .0000 &   .0000 &   .0101     
\end{tabular} \right) \hspace*{0.2cm}
\begin{array}{r}
 k = 1 \\ 2 \\ 3 \\ 4 \\ 5 \\  6 
\end{array} \hspace*{0.3cm}
\left( \begin{array}{r}
   979.331\\
  1050.701\\
  1050.701\\
  1052.701\\
  1102.331\\
  1102.397
\end{array} \right)
 \label{diagdw}
\eeq  
where the $6\times 3$ matrices on the left and right of the 
vertical bar are $\Gm_{DL} $ and $\Gm_{DR}$. The differences in 
the elements (1,1), (2,2), (4,4), (5,5) in ${\wti M}_U$ and 
${\wti M}_D$ are due to the different D--terms present in the 
two sectors. The off--diagonal elements (1,3), (2,3) in 
${\wti M}_D$ are induced by the non-diagonal up quark 
yukawa couplings during the evolution from $X$ to $Q_0$. The 
same elements in ${\wti M}_U$ sum also the supersymmetry
non--violating terms induced by up quark yukawa couplings.


\newpage
\small 
\begin{thebibliography}{99}

\bibitem{DNNS}
 M.~Dine, A.E.~Nelson, and Y.~Shirman, \prd{51}{95}{1362};  \\
 M. Dine, A.E. Nelson, Y. Nir, and Y. Shirman, \prd{53}{96}{2658}

\bibitem{BABU}
  K.S. Babu, C. Kolda, and F. Wilczek, \prl{77}{96}{3070}

\bibitem{DNS} 
 M. Dine, Y. Nir, and Y. Shirman, \prd{55}{97}{1501}

\bibitem{DIM}
 S.~Dimopoulos, S.~Thomas, and J.D.~Wells, \hph{96}{09}{434} 

\bibitem{BAGG}
 J.A.~Bagger, K.~Matchev, D.M.~Pierce, and R.~Zhang, \prd{55}{97}{xxx}, 
 \hph{96}{09}{444} 

\bibitem{SMARTIN}
  S. Martin, \prd{55}{97}{xxx}, \hph{96}{08}{224}

\bibitem{DGP}
  S. Dimopoulos, G.F. Giudice, and A. Pomarol, \hph{96}{07}{225}

\bibitem{NIR}
 N.~Polonsky, \prd{54}{96}{4537} 

\bibitem{BBMR}
  S.~Bertolini, F.M.~Borzumati, A.~Masiero, and 
                               G.~Ridolfi,~\npb{353}{91}{591}
\bibitem{CAPOWA}
 M.~Carena, S.~Pokorski, and C.~Wagner, \npb{426}{94}{269}

\bibitem{RASA} 
  R.~Rattazzi and U.~Sarid, \prd{53}{96}{1553}   

\bibitem{MAVAU}
 S.~Martin and M.~Vaughn, \prd{50}{94}{2282}

\bibitem{GRZ}
 G.~Gamberini, G.~Ridolfi, and F.~Zwirner, \npb{331}{90}{331}

\bibitem{CASAS}
 J.A.~Casas, A.~Lleyda, and C.~Munoz, \npb{471}{96}{3} 

\bibitem{HAKA}
 H.E.~Haber and G.L.~Kane, \prep{117}{85}{75}

\bibitem{CDFPHPHTH} 
 S.~Dimopoulos, M.~Dine, S.~Raby, and S.~Thomas, 
        \prl{76}{96}{3494} \\
 S.~Ambrosanio, G.~Kane, G.~Kribs, S.~Martin, and S.~Mrenna,
        \prl{76}{96}{3498} 

\bibitem{GMSBbsg}
 H.~Baer, M.~Brhlik, C.~Chen, and X.~Tata, hep--ph \#9610358 \\
 N.G.~Deshpande, B.~Dutta, and S.~Oh, \hph{96}{11}{443}


\bibitem{BMMP}
 A.~J.~Buras, M.~Misiak, M.~M\"unz, and S.~Pokorski, 
                                 \npb{424}{94}{374} 

\bibitem{BDN}
 F.M.~Borzumati, M.~Drees, and M.M.~Nojiri, \prd{51}{95}{27}

\bibitem{IO}
 F.M.~Borzumati, \zpc{63}{94}{291}

\bibitem{CLEO} 
 M.S.~Alam, CLEO Collaboration, \prl{74}{95}{2885}

\bibitem{CDFPHPH}
 D.~Toback, CDF Collaboration, DPF96 proceedings \\
 K.~Wyatt Merrit, ``Searches for New Physics at the Tevatron'', 
  DPF96, hep-ex \#9701009 

\bibitem{GUHA}
 J.F.~Gunion and H.E.~Haber, \npb{272}{86}{1}

\bibitem{ZWIRN1}
 J.~Ellis, G.~Ridolfi, and F.~Zwirner, \plb{262}{91}{477} 

\bibitem{MANUEL}
 M.~Drees and M.~Nojiri, \prd{45}{92}{2482}

\bibitem{ZWIRN2}
 A.~Brignole, J.~Ellis, G.~Ridolfi, and F.~Zwirner, 
                                     \plb{271}{91}{123}

\bibitem{VIENNA}
 H.~Eberl and W.~Majerotto, 
 Vienna University preprint, HEPHY-PUB 595/93 (1993), unpublished.

\bibitem{DENNIS}
 D.~Comelli, M.~Pietroni, and A.~Riotto, \prd{50}{94}{7703}

\end{thebibliography}
\end{document}